\documentclass{aa}
\usepackage[colorlinks=true, linkcolor=blue, citecolor=blue, filecolor=blue, urlcolor=blue]{hyperref}
\usepackage{graphicx}
\usepackage{txfonts}
\usepackage{color}
\usepackage{xspace}
\usepackage{float}
\usepackage{url}
\usepackage{amsmath}
\usepackage{amssymb}
\usepackage{xcolor}
\newcommand{\g}{\ensuremath{\gamma}\xspace}
\newcommand{\hd}{H$_2$\xspace}
\newcommand{\hi}{H\textsc{i}\xspace}
\newcommand{\nh}{\ensuremath{N_{\rm{H}}}\xspace}
\newcommand{\nhi}{\ensuremath{N_{\rm{HI}}}\xspace}
\newcommand{\nhp}{\ensuremath{N_{\rm{H}^+}}\xspace}

\newcommand{\nhdnm}{\ensuremath{N_{\rm{H}_{\rm{DNM}}}}\xspace}
\newcommand{\opa}{$\tau_{353}/N_{\rm{H}}$\xspace}
\newcommand{\qlis}{\ensuremath{q_{\rm{LIS}}}\xspace}
\newcommand{\taunu}{$\tau_{353}$\xspace}
\newcommand{\ts}{\ensuremath{T_{\rm{S}}}\xspace}
\newcommand{\percc}{cm$^{-3}$\xspace}
\newcommand{\persqcm}{cm$^{-2}$\xspace}
\newcommand{\cmsqs}{cm$^2$\,s$^{-1}$\xspace}
\newcommand{\halpha}{H$_\alpha$\xspace}
\newcommand{\bsky}{$B_{\rm sky}$\xspace}

\def\Fermi{\textit{Fermi}\xspace}
\def\Planck{\textit{Planck}\xspace}
\def\IRAS{\textit{IRAS}\xspace}

\begin{document}

   \title{Cosmic-ray diffusion in two local filamentary clouds}
\author{
F.R.~Kamal~Youssef \inst{1} \and 
I.~A.~Grenier \inst{1} 
}
\authorrunning{LAT collaboration}

\institute{
Universit\'{e} Paris Cit\'{e} and Universit\'{e} Paris-Saclay, CEA, CNRS, AIM, F-91190 Gif-sur-Yvette, France\\ 
\email{isabelle.grenier@cea.fr},  
%\url{https://orcid.org/0000-0003-3274-674X}
}

   %\date{\today}
   \date{Submitted October 17, 2023 / accepted February 14, 2024}
   
\abstract
% Context
{Hadronic interactions between cosmic rays (CRs) and interstellar gas have been probed in \g rays across the Galaxy. A fairly uniform CR distribution is observed up to a few hundred parsecs from the Sun, except in the Eridu cloud, which shows an unexplained 30--50\% deficit in GeV to TeV CR flux.}
% Aims
{To explore the origin of this deficit, we studied the Reticulum cloud, which shares notable traits with Eridu: a comparable distance in the low-density region of the Local Valley and a filamentary structure of atomic hydrogen extending along a bundle of ordered magnetic-field lines that are steeply inclined to the Galactic plane. 
%We have compared the \g-ray and insterstellar properties of the two clouds at the same parsec scale.
} 
% Methods
{We measured the \g-ray emissivity per gas nucleon in the Reticulum cloud in the 0.16 to 63 GeV energy band using 14 years of \Fermi-LAT data. We also derived interstellar properties that are important for CR propagation in both the Eridu and Reticulum clouds, at the same parsec scale. 
%They include the gas volume density, the magnetic to thermal pressure ratio, the amplitude of turbulent perturbations in gas velocity and in magnetic-field orientation, the turbulence compressibility, and the steady-state diffusion coefficient of the particles along the magnetic field in the self-confinement scenario.
} 
% Results
{The \g-ray emissivity in the Reticulum cloud 
is fully consistent with the average spectrum measured in the solar neighbourhood, but this emissivity, and therefore the CR flux, is $1.57 \pm 0.09$ times larger than in Eridu
%its Eridu homologue 
across the whole energy band. The difference cannot be attributed to uncertainties in gas mass. Nevertheless, we find that the two clouds are similar in many respects: both have magnetic-field strengths of a few micro-Gauss in the plane of the sky; both are %(only 60\% larger in Eridu); 
in approximate equilibrium between magnetic and thermal pressures; they have similar turbulent velocities and sonic Mach numbers;
%of a few kilometres per second (hardly twice as large as in the more diffuse Eridu than in the more compact Reticulum cloud), with similar sonic Mach numbers; and a similar
and both show magnetic-field regularity with a dispersion in orientation lower than 10$^\circ$-15$^\circ$ over large zones.
%parts of each filament. 
%The gas density estimates and the detection of dark neutral gas at the \hi to \hd transition in Reticulum, but not in Eridu, indicate that
%The atomic gas in Reticulum is typical of the dense cold neutral medium whereas the gas in Eridu is more likely in the unstable lukewarm phase. 
The gas in Reticulum is colder and denser than in Eridu,
%We find, however, 
but we find similar parallel diffusion coefficients around a few times $10^{28}$~\cmsqs in both clouds 
%in the scenario where 
if CRs above 1 GV in rigidity diffuse on resonant, self-excited Alfvén waves that are damped by ion--neutral interactions. }
% Conclusions
{The loss of CRs in Eridu remains unexplained, but these two clouds provide important test cases to further study how magnetic turbulence, line tangling, and ion--neutral damping regulate CR diffusion in the dominant  gas phase of the interstellar medium.}

%                Galaxy: solar neighbourhood --
%                ISM: dust --
\titlerunning{cosmic rays in local gas filaments}

\maketitle

%------------Partie 1

\section{Introduction}

The Sun is located in a valley of low gas density surrounded by two walls of dense molecular clouds \citep[see Fig. 2 of][]{Leike20}. Part of the valley, out to ${\sim}$150 pc or ${\sim}$250 pc depending on the direction, corresponds to the Local Bubble \citep[see Fig. 1 of][]{Pelgrims20}, but the elongated valley extends further away in opposite directions roughly parallel to the local spiral arm. 
Interactions of cosmic rays (CRs) with the interstellar gas and radiation fields present inside and around the valley produce \g rays that have been recorded at energies above 30 MeV by the Large Area Telescope (LAT) on board the \Fermi Gamma-Ray Space Telescope. 
As the hadronic contribution to this \g-ray intensity integrates the product of the CR and gas volume densities along the lines of sight, the \g-ray map of nearby clouds can be used to probe the CR flux in different locations and in different types of environment \citep{Lebrun82}. This method requires an accurate measurement of the gas mass in a cloud. 

In the warm atomic phase of clouds, often referred to as the warm neutral medium (WNM), one can directly infer the gas mass from observations of the \hi 21~cm lines \citep[e.g.][]{Dickey00,Heiles03}. In the denser, cold atomic phase of the clouds, where the \hi brightness temperature probes the excitation (spin) temperature of the gas rather than the column density, the lines must be corrected for partial self-absorption in order to estimate the gas mass \citep[][and references therein]{Murray15}. This cold atomic phase, referred to as the cold neutral medium (CNM), generally contributes a fraction between 5\%\ and 28\% of the total atomic mass \citep{Murray18,Murray20}. %\cite{kalberla10_gassHI} ; \cite{peek11_galfaHI}. 
\hd molecules radiate inefficiently in the cold conditions of molecular clouds, so we must rely on observations of surrogate lines, such as the rotational lines of $^{12}$CO and its isotopologues, in order to quantify the \hd mass. This is only possible where the abundance and excitation temperature of CO molecules are sufficient. Environmental variations of the $X_{\rm CO}$ conversion factor between the $^{12}$CO(1-0) line intensity and the \hd column density have been studied theoretically and observationally \citep[][and references therein]{Bolatto13,Gong18,Remy17}, but the variations are too large and too uncertain to reliably use the CO data to measure the CR flux in the dense molecular phase of clouds.

Complementary information can be obtained from the thermal emission of large dust grains mixed with the gas. The emission is observed at submillimetre and infrared wavelengths \citep{Planck14XI,Planck14XVII}. Dust grains and CRs are both tracers of the total gas in all chemical and thermodynamical forms \citep{Grenier05,Planck15_Cham}. The \g rays reveal the gas mass convolved by the ambient CR flux, whereas the dust optical depth shows the gas mass convolved by the ambient dust-to-gas mass ratio and by the ambient grain emissivity. Both the CR and dust properties vary from place to place, but the spatial correlation between the two tracers in a given cloud can be used to trace the dark gas that escapes \hi and CO line observations \citep{Grenier05}. This dark neutral medium (DNM) accumulates at the \hi--\hd transition in the clouds \citep{Liszt19}. It is not seen in \hi or CO emission because of self-absorption in the dense \hi clumps and because CO molecules are efficiently photo-dissociated or too weakly excited in the diffuse \hd gas. 

Given the uncertainty in gas mass estimates in the dark and molecular phases of clouds, one can only infer the CR spectrum with adequate precision in the atomic phase of the interstellar medium (ISM). The derivation of the \g-ray emissivity, $q_{\rm HI}$, per gas nucleon in the atomic gas relies on the spatial correlation between the \g-ray intensity and \hi column density (\nhi), but the analysis must include maps of the other gas phases in order to avoid biasing the \hi emissivity spectrum \citep{Planck15_Cham}. These spectra are typically obtained at energies above a few hundred MeV in order to take advantage of the better angular resolution of the LAT in order to resolve the clouds and their different phases. 

Such emissivity measurements have been performed in different clouds of the local ISM \citep[e.g.][]{Ackermann10_Cep,Ackermann12_Orion,Planck15_Cham,Casandjian15,Tibaldo15,Remy17,Joubaud19} and across the Milky Way \citep[e.g.][]{Ackermann13_3rdQ,Acero16_diffuse}. The local emissivity spectra appear to be often consistent with the local average, $q_{\rm LIS}(E)$, which has been obtained for the whole atomic gas seen at Galactic latitudes ranging between 7$^\circ$ and 70$^\circ$ \citep{Casandjian15}. This consistency implies that the CR flux is rather uniform within a few hundred parsecs of the Sun \citep{Grenier15}.

At variance with this uniformity, a significantly lower \g-ray emissivity spectrum has been found in a nearby cloud named Eridu.  The spectrum is 34\% lower than the \qlis average  \citep{Joubaud20}. The deviation is statistically significant (14 $\sigma$) and there is no confusion with other clouds along the lines of sight. The deviation cannot be attributed to uncertainties in the \hi mass because the minimum amount of gas inferred from the \hi lines (optically thin case) gives an upper limit to the  emissivity that is still 25\% below \qlis. More CRs would outshine the observed \g rays. The \g-ray analysis has probed CRs with energies above several GeV. Their penetration in the cloud is likely unhampered given the modest column densities (1--5 $\times 10^{20}$~\persqcm) and low volume densities (around 7~\percc) of the diffuse gas \citep{Joubaud19}. The Eridu cloud therefore appears to be pervaded by a ${\sim}30$\% lower CR flux, but with the same particle energy distribution as in the local ISM.

\begin{figure}[!ht]
  \centering     
  \includegraphics[scale=0.5]{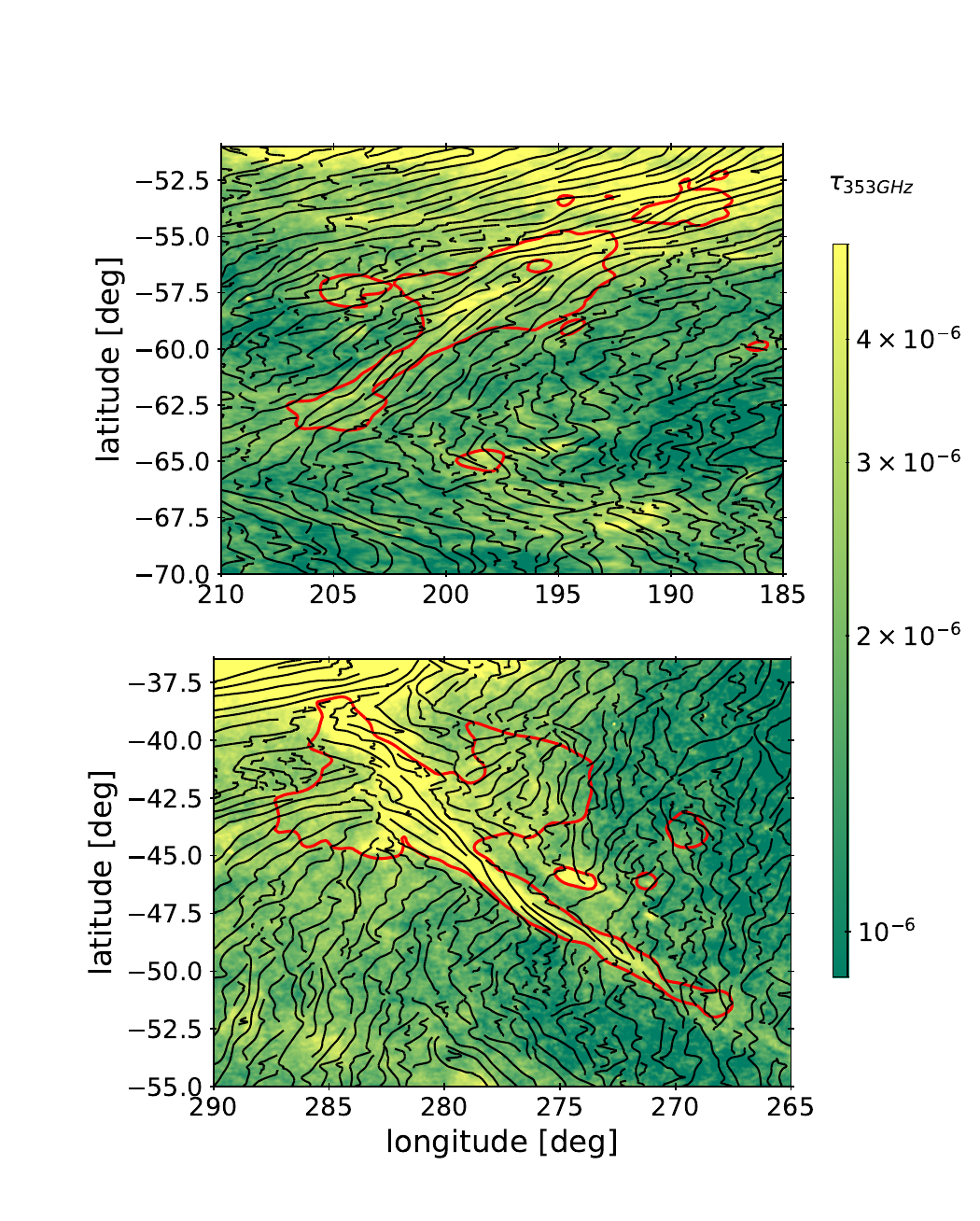}
  \caption{Dust optical depth $\tau_{\rm 353}$ at 353~GHz in Galactic coordinates towards the Eridu (top) and Reticulum (bottom) filaments. The black lines give the orientation of the plane-of-the-sky $B_{\rm sky}$ magnetic field inferred from the dust polarisation observations at 353~GHz and at an angular resolution of 14\arcmin~(FWHM). The red contours delineate the filaments at \hi gas column densities of $3 \times 10^{20}$ cm$^{-2}$ and $1.2 \times 10^{20}$ cm$^{-2}$, respectively. }  
  \label{fig:Bskymaps}
\end{figure}

The Eridu cloud is a diffuse, atomic, and filamentary cirrus. It is located outside the edge of the Orion-Eridanus superbubble, at high Galactic latitudes of around 50$^\circ$ and at an altitude of about 200--250~pc below the Galactic plane \citep{Joubaud19}. Figure \ref{fig:Bskymaps} shows that the filament is inclined with respect to the Galactic plane and the dust polarisation data at 353~GHz from \Planck \citep{Planck15XIX} show that the filament extends along a bundle of well-ordered magnetic-field lines, pointing within 45$^\circ$ of the southern Galactic pole. 

Magnetic loops are pushed out of the Galactic disc by gas fountains. In order to investigate their impact on CR transport to the halo, we searched for another filament with gaseous and magnetic characteristics comparable to those of Eridu. 
The Reticulum cloud is also an atomic and filamentary cloud that lies at the edge of the Local Valley, at a distance we estimate to be about 240~pc (Appendix \ref{sec:dist}). Figure \ref{fig:Bskymaps} shows that, like Eridu, it is largely inclined with respect to the Galactic plane. It points towards the southern Galactic halo and has well-ordered magnetic field lines aligned with the filament axis. We therefore aim to measure the \g-ray emissivity spectrum per gas nucleon in the Reticulum filament and to compare it with the average in the local ISM and with the emissivity spectrum in the Eridu cloud.

The paper is structured as follows: The data are presented in Sect. \ref{sec:data} and the models are described in Sect. \ref{sec:mod}. The results and the detection of dark gas are detailed in Sect. \ref{sec:results}. The CR content of the clouds is discussed in Sects. \ref{sec:densities} to \ref{sec:Bsky}.  

\begin{figure}[ht]
  \centering
  \includegraphics[width=\hsize]{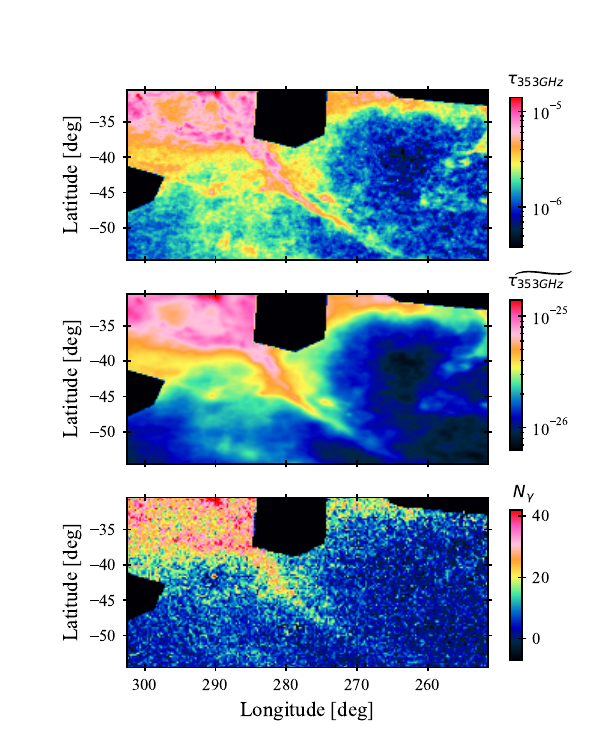}  \caption{Dust optical depth measured at 353~GHz from the \Planck and IRAS data and displayed on a 10\farcm2$\times$7\farcm5 grid at the angular resolution of \Planck (top panel) and of \textit{Fermi}-LAT (middle). The bottom panel shows the \g-ray photon counts produced by CR interactions with gas and recorded by \textit{Fermi}-LAT in 0\fdg25$\times$0\fdg25 bins in the 0.16--63 GeV energy band (other emissions of non-gaseous origin are subtracted). Three zones are excluded from the analysis region : the SMC on the left, LMC 
  in the middle, and the edge of the Gum nebula in the upper right. The Reticulum cloud is the long filament crossing the region below the LMC.}  
  \label{fig:gamDust}
\end{figure}

\begin{figure*}[ht]
\includegraphics[width=\hsize]{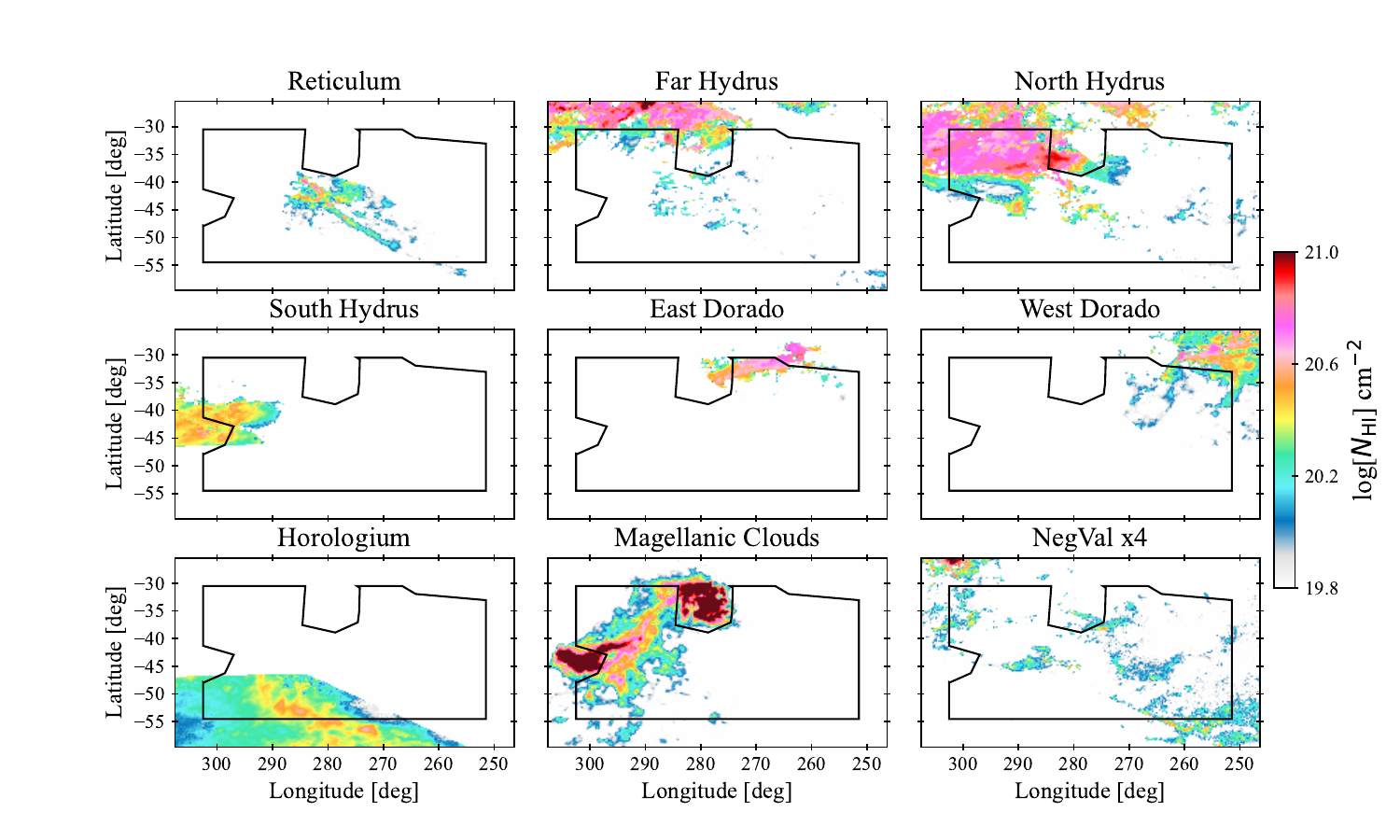}
\caption{\nhi column densities in the Reticulum (Ret), Far Hydrus (HyiF), North Hydrus (HyiN), South Hydrus (HyiS), East Dorado (DorE), West Dorado (DorW), Horologium (Hor), and the Magellanic clouds for optically thin \hi. The faint column densities in the negative-velocity map (NegVel) are multiplied by 4 for display. The maps include a 5$^\circ$-wide peripheral band around the analysis region (black perimeter) to account for the wings of the \Fermi-LAT PSF in the \g-ray model.} 
\label{fig:NHI}
\end{figure*}

\section{Data}
\label{sec:data}

We selected an analysis region around the Reticulum filament extending from 251\fdg5 to 302\fdg5 in Galactic longitude and from $-$54\fdg5 to $-$30\fdg5 in Galactic latitude, as shown in Fig. \ref{fig:gamDust}. We masked regions with large column densities from the Large Magellanic Cloud (LMC) and Small Magellanic Cloud (SMC), and from the edge of the Gum Nebula.
 
\subsection{Dust data}

In order to trace the dust column density, we used the all-sky map of the dust optical depth, $\tau_{353}$, at 353~GHz (\citet{Planck14XI}, updated by \citet{Irfan19}). It was constructed at an angular resolution of 5\arcmin~by modelling the intensity, $I_{\nu}$, of the dust thermal emission recorded in frequency $\nu$ by \Planck and the Infrared Astronomical Satellite (\IRAS). The dust optical depth relates to the modified black-body intensity $B_\nu(T_d)$ for a dust temperature $T_d$ and spectral index, $\beta$, according to 
$I_\nu=\tau_{353} \, B_\nu(T_d) (\nu/353~\textrm{GHz})^\beta$. 
The dust opacity at 353~GHz is defined for a hydrogen column density $N_{\rm H}$ as the ratio \opa.

Figure \ref{fig:gamDust} shows the distribution of the dust optical depth in the analysis region at the original \Planck resolution in the upper panel and at the angular resolution of \Fermi-LAT for the 0.16--63~GeV band in the middle one. 

\subsection{Gamma-ray data}

We used 14 years of \Fermi-LAT Pass 8 P8R3 
survey data for energies between 0.16 and 63~GeV \citep{bruel18} and we analysed the data in eight energy bands that are bounded by 10$^{2.2}$, 10$^{2.4}$, 10$^{2.6}$, 10$^{2.8}$, 10$^{3.0}$, 10$^{3.2}$, 10$^{3.6}$, 10$^{4.0}$, and 10$^{4.8}$~MeV. 

The \Fermi-LAT point spread function (PSF) strongly varies with energy. In order to retain sufficient photon statistics in this faint region and to preserve a good contrast in photon counts across small cloud structures, we combined photons of different PSF types in each energy band. Photons are labelled from the worst (PSF0) to best (PSF3) quality in angle reconstruction. We used PSF 3 in the three lowest-energy bands, PSF 2 and 3 in the next band, PSF 1 to 3 in the next two bands, and PSF 0 to 3 in the two highest energy bands. The 68\% containment radius of the resulting PSF varies from 0\fdg9 above 630 MeV to 1\fdg4 at 250 MeV and to 2\fdg1 at 160 MeV. The resulting photon map is shown in the lower panel of Fig. \ref{fig:gamDust} where we subtracted the point sources and the very smooth isotropic and inverse-Compton (IC) components to display the \g rays produced in the clouds.

In order to reduce to less than 10\% the contamination of the photon data by residual CRs and by Earth atmospheric \g rays, we applied tight selection criteria such as the SOURCE class and photon arrival directions within an energy-dependent and PSF-type-dependent angle varying from 90$^\circ$ to 105$^\circ$ from the Earth zenith, namely 90$^\circ$ in the first energy band, 100$^\circ$ in the second, and 105$^\circ$ at higher energies. We used the associated instrument response functions P8R3\_SOURCE\_V3.

We modelled the interstellar emission in a broader region than the analysis one in order to take into account the extent of the \Fermi-LAT PSF. The 5$^\circ$-wide peripheral band around the analysis perimeter corresponds to the PSF $>$95\% containment radius for our selections. It is displayed in Fig. \ref{fig:NHI}. 

The positions and flux spectra of the \g-ray sources in the field were provided by the \Fermi-LAT 4FGL-DR3 %\footnote{https://fermi.gsfc.nasa.gov/ssc/data/access/lat/12yr\_catalog/} 
12-year source catalogue \citep{Abdollahi22_DR3}. The analysis region and the 5$^\circ$-wide peripheral band contain 220 point sources. Their flux spectra were calculated with the spectral characteristics given in the catalogue.

The observation also includes IC \g rays produced by CR electrons up-scattering the interstellar radiation field in the Milky Way. 
This was modelled by GALPROP\footnote{\url{http://galprop.stanford.edu}} and we used the LRYusifovXCO5z6R30\_Ts150\_Rs8 model version \citep{ackermann12}. The isotropic spectrum for the extragalactic and residual instrumental backgrounds is given by the \Fermi-LAT science support centre\footnote{\url{http://fermi.gsfc.nasa.gov/ssc/data/access/lat/BackgroundModels.html}}. We used the isotropic spectrum provided for each PSF type of the SOURCE class.

\subsection{Gas data}
\label{sec:HIdata}

In order to map the neutral atomic hydrogen (\hi) in the region, we used the 21-cm Galactic All Sky Survey GASS III data \citep{Kalberla15}. The survey has an angular resolution of 16\farcm2, a velocity resolution of 0.82~km/s in the local standard of rest, and an rms sensitivity of 57~mK. Lines in the analysis region span velocities from $-$125 to +495~km/s. 

Following the method developed by \cite{Planck15_Cham}, we decomposed the \hi spectra into sets of individual lines with Pseudo-Voigt profiles. Based on the ($l$, $b$, $v$) distribution of the fitted lines, we identified seven nearby cloud complexes that are coherent in velocity and in position across the region. We named these clouds according to the constellation in their direction, namely Reticulum (Ret), Far Hydrus (HyiF), North Hydrus (HyiN), South Hydrus (HyiS), East Dorado (DorE), West Dorado (DorW), and Horologium (Hor).
They are displayed in Fig. \ref{fig:NHI} and they span velocities from 3.3 to 20.6 km/s, -15.7 to -1.6 km/s, -0.8 to 13.2 km/s, -14.8 to 2.5 km/s, 2.5 to 4.9 km/s, 5.8 to 33.8 km/s, and -14.8 to 5.8 km/s, respectively. These clouds are local. We gathered lines from the Magellanic clouds in a single map. We added a ninth component gathering the very weak lines found at velocities below $-$8~km/s. 
 
 We produced \nhi column-density maps for each complex by integrating the fitted lines that have a central velocity within the chosen velocity interval for each ($l$, $b$) direction. This method corrects for potential line spill-over from one velocity interval (i.e. cloud) to the next. The line fits produce small residuals between the modelled and observed \hi spectra. In order to preserve the total \hi intensity, we distributed these residuals among the fitted lines according to their relative strength in each velocity channel. 

In order to correct the \hi intensity for potential self-absorption, we produced the \nhi maps for a set of uniform spin temperatures, \ts, of 100, 125, 150, 200, 300, 400, 500, 600, 700, and 800 K and also for the optically thin case. \citet{Nguyen18} have shown that a simple isothermal correction of the emission spectra with a uniform spin temperature can well reproduce the more precise \nhi column densities that are inferred from the combination of emission and absorption spectra. Given the wide range of potential spin temperatures we could not test individual temperatures for each cloud, but we used the same temperature for all clouds and we let it vary when modelling the dust and \g-ray data.

The nearby clouds in the region are mainly composed of atomic gas. We looked for CO(1-0) emission at 115~GHz towards them, but did not find published observations in that area. We checked that \Planck detected no CO(1-0) intensity up to 1 K km/s  in the region \citep{Planck14XIII}.

\subsection{Cloud distances}
\label{sec:cloudDist}
\citet{Leike20} used the stellar data from the \textit{Gaia}-DR2, ALLWISE, PANSTARRS, and 2MASS surveys to model the dust extinction and to reconstruct the 3D map of dust clouds within ${\sim}400$~pc from the Sun with a spatial resolution of 2~pc. Distances to compact clouds and their dust densities are expected to be well constrained by the precise locations of the stars. In Figs. \ref{fig:dist_Ret} and \ref{fig:dist_Eri} we display cuts of the 3D dust distribution in 20~pc wide intervals, spanning distances from 150~pc to about 350~pc. The shapes and locations of the clouds isolated in the \hi data in Fig. \ref{fig:NHI} can be roughly recognised in the dust distribution. We also probed the dust distribution along lines of sight exhibiting large gas column densities in the different clouds (see Fig. \ref{fig:distprof_Ret}). From the dust and gas correspondence we infer average distances of $240\pm30$~pc, $220\pm30$~pc, $220\pm30$~pc, $170\pm30$~pc, $310\pm30$~pc, $340\pm30$~pc, and $280\pm30$~pc to the Reticulum, North Hydrus, South Hydrus, Far Hydrus, East Dorado, West Dorado, and Horologium clouds, respectively. The $\pm 30$ pc uncertainty reflects the size of dust overdensities that can be isolated in the 3D dust cube.

We applied the same method to revisit the distance to the Eridu cloud. The rim of the Orion-Eridanus superbubble is clearly detected at a distance around 200 pc in the dust 3D data. The Eridu cloud lies beyond this edge. Figure \ref{fig:dist_Eri} shows a second dust front at a distance of $300 \pm 30$ pc with an inclination and extent in the sky that is compatible with that of the elongated Eridu cloud seen in \hi. This distance places the cloud at an altitude of 260 pc below the Galactic plane, near the edge of the ${\sim}220$~pc thick \hi layer of the local ISM \citep{Nakanishi16}. A much larger distance would be disfavoured. The \hi line velocities of the cloud are also typical of the local ISM and not of intermediate- or high-velocity clouds outside the disc. Extending the 3D dust analysis to larger distances than available in the \cite{Leike20} cube would help confirm the distance to Eridu. 

\section{Models and analyses} \label{sec:mod}

Figure \ref{fig:gamDust} shows that the dust optical depth and the \g-ray intensity produced by CRs in the different clouds are strongly correlated. The dust and \g rays both trace, to first order, the total gas column density $N_{\rm H}$. 
In order to detect neutral gas not observed in the \hi data, we used the fact that it is permeated by both CRs and dust grains. We therefore extracted \g-ray and dust residuals between the observation and the model expected from the \nhi distribution. We then used the spatial correlation between the positive residuals found in dust and in \g rays to extract the additional gas column densities (see Sect \ref{sec:iter}). This additional gas corresponds to the DNM at the \hi--\hd transition where the dense \hi becomes optically thick to 21-cm line emission and CO emission is absent or too faint to trace the \hd gas. The analysis results yield small amounts of dark gas in the Reticulum and North Hydrus clouds (see Sect. \ref{sec:DNM}).

\subsection{Dust model}

Provided that the dust-to-gas mass ratio and the mass emission coefficient of the grains are uniform (but not necessarily the same) in each gas phase of a particular cloud, the dust optical depth $\tau_{353}(l,b)$ can be modelled as a linear combination of the different gas column densities (\nhi and \nhdnm), each with free $y$ normalisation to be fitted to the data, as in  \cite{Planck15_Cham}. A free isotropic term, $y_{\rm iso}$, was added to the model to account for the residual noise and the uncertainty in the zero level of the dust maps \citep{Planck14XI,Casandjian22}.\\
The $\tau_{353}(l,b)$ model can be expressed as :
\begin{equation}
\tau_{353}(l,b)= \sum_{i=1}^9 y_{\rm{HI},i} N_{\rm{HI},i}(l,b) + \sum_{i=1}^2 y_{\rm{DNM},i} N_{\rm{H}~\gamma}^{\rm{DNM},i}(l,b)
+ y_{\rm{iso}} ,
\label{eq:modDust}
\end{equation}
where $N_{\rm{HI},i}(l,b)$ denotes the nine \nhi maps depicted in Fig. \ref{fig:NHI} and $N_{\rm H~\gamma}^{\rm{DNM},i}(l,b)$ stands for the column density in the two DNM components constructed from the \g-ray data (see Sect. \ref{sec:iter}). We degraded the optical-depth map from its original resolution of 5\arcmin~ down to 16\farcm2~ to match that of the GASS \hi survey. 

The $y_{\rm{HI},i}$ coefficients give the average values of the dust \opa opacity in the different clouds. 
%\hi maps. 
The $y_{\rm{DNM},i}$ parameters can probe opacity changes in the denser DNM phases.
They were fitted to the data using a least-square $\chi^{2}$ minimisation. We expect the uncertainties of the model to exceed those of the dust map because of our assumption of uniform grain distributions through the clouds and because of the limited capability of the \hi data to trace the total gas. As we cannot precisely estimate the model uncertainties, we set a fractional error of 11\% that yields a reduced $\chi^{2}$ value of one in the dust fit. This fraction is slightly larger than the 1\% to 9\% measurement uncertainties in the $\tau_{353}$ values across this region.

\subsection{Gamma-ray model}
\label{sec:gmodel}

\begin{figure*}%[!ht]
  \centering 
  \includegraphics[width=\hsize]{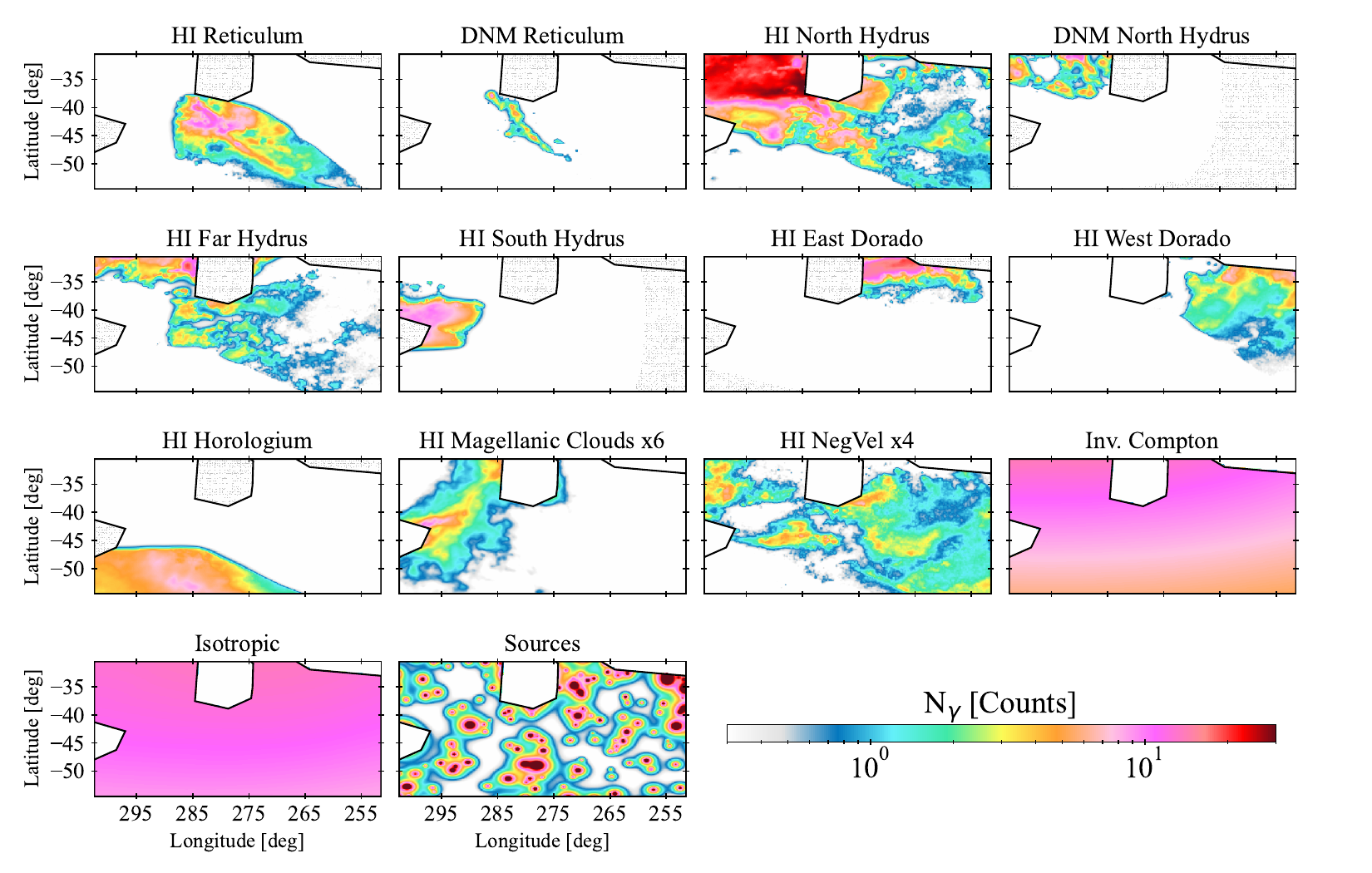}
  \caption{ Photon yields on a 
0\fdg25 pixel grid from the main components of the \g-ray model in the 0.16--63 GeV band. The yields come from the \hi and DNM column densities derived for the optically thin case in the different clouds, and from the isotropic background, Galactic IC emission, and \g-ray point sources.}
  \label{fig:gamMod}
\end{figure*}

Earlier studies have indicated that the Galactic CRs radiating at 0.1–100~GeV have diffusion lengths much larger than typical cloud dimensions and that they propagate through the different gas phases \citep{Grenier15}. The observed interstellar \g-ray emission can therefore be modelled, to first order, by a linear combination of the gas column densities present in the different clouds seen along the lines of sight, in both the \hi and DNM gas phases. The model also includes a contribution from the Galactic IC intensity, $I_{\rm IC}(l,b,E)$, point sources of non-interstellar origin with individual flux spectra $S_{j}(E)$, and an isotropic intensity, $I_{\rm iso}(E)$, accounting for the extragalactic \g-ray background and for residual CRs misclassified as \g rays.
We verified that the soft emission from the Earth limb is not detected in the present energy range for the choice of maximum zenith angles.

The \g-ray intensity, $I(l,b,E)$, in \g cm$^{-2}$ s$^{-1}$ sr$^{-1}$ MeV$^{-1}$ in each $(l,b)$ direction, is modelled in each energy band E as : 
\begin{eqnarray}
      I(l,b,E) &=& q_{\rm LIS}(E) \,\times \, \bigr[\, \sum_{i=1}^9 q_{\rm{HI},i}(E) \, N_{\rm{HI},i}(l,b) \, \,\,\bigr] \nonumber\\
      &+& q_{\rm LIS}(E) \,\times \, \bigr[\, \sum_{i=1}^2 q_{\rm{DNM},i}(E) \, \tau_{353}^{\rm{DNM},i}(l,b) \, \,\,\bigr] \nonumber \\
      &+&\, q_{\rm{IC}}(E) \, I_{\rm IC}(l,b,E) \,\, + \,\, q_{\rm iso}(E) \, I_{\rm iso}(E) \nonumber\\
       &+&\, \sum_j q_{S_j}(E) \, S_j(E) \, \delta(l-l_j,b-b_j) ,
\label{eq:modGam}
\end{eqnarray} 
where $N_{\rm{HI},i}(l,b)$ denotes the \hi column density map of each cloud and the extraction of the dust DNM $\tau_{353}^{\rm{DNM},i}$ maps is explained in Sect. \ref{sec:iter}. The $q$ coefficients of the model are determined from fits to the \Fermi-LAT data.

The $q_{\rm{HI},i}$, $q_{\rm IC}$, and $q_{\rm iso}$ parameters are simple normalisation factors to account for possible deviations from the input spectra taken for the local gas \g-ray emissivity $q_{\rm LIS}(E)$ and for the isotropic and IC intensities.
The input \qlis spectrum was based on the correlation found between the \g radiation and the \nhi column densities derived from the Leiden-Argentine-Bonn (LAB) survey for a spin temperature of 140~K, at latitudes $10^\circ \leq |b| \leq 70^\circ$ \citep{Casandjian15}. The $q_{\rm{HI},i}$ parameter in the model can therefore compensate for differences in the \hi data (calibration, angular resolution, spin temperature) and for cloud-to-cloud variations in CR flux. Such differences will affect the normalisation equally in all energy bands whereas a change in CR penetration inside a specific cloud will show as an energy-dependent correction, primarily as a loss at energies below one GeV \citep{Schlickeiser16}. 

The source flux spectra $S_j(E)$ were calculated with the spectral characteristics given in the 4FGL-DR3 catalogue. The $q_{S_j}(E)$ normalisation factors in the model allow for possible changes due to the longer exposure used here and to the use of a different interstellar background for source detection in the 4FGL-DR3 catalogue. 
The $q_{S_j}(E)$ normalisation factors of the sources inside the analysis region or in the peripheral band were fitted individually, except for those inside the SMC and LMC regions and for those found in the lowest dust zones ($\tau_{353} < 3 \times 10^{-6}$), which were treated collectively in three separate sets with a single normalisation factor per set.  

In order to compare the model with the LAT photon data in the different energy bands, we multiplied each component map by the energy-dependent exposure and we convolved it with the energy-dependent PSF and energy resolution. The convolution with the response functions was done for each PSF event type independently in small energy bands. We also calculated the isotropic map for each PSF event type in the same small energy bands. We then summed the photon count maps in energy and for the different combinations of PSF types to get the modelled count maps in the eight energy bands of the analysis for the same photon selection as in the data. 

Figure \ref{fig:gamMod} gives the photon yields that were obtained in the entire energy band. 
% from the best fit
It shows that the emission originating from the gas can be distinguished from the much smoother emission backgrounds and that the LAT angular resolution is sufficient to separate the various clouds, as well as the different gas phases within the clouds. The distinct spatial distributions of the clouds and of their phases allow rather independent measurements of their individual \g-ray emissivities. On the contrary, we do not expect a firm separation of the isotropic and IC spectra across this region. We used a binned maximum-likelihood with Poisson statistics to fit the model coefficients $q$ to the LAT data in each of the eight energy bands and in the overall one.

\subsection{Analysis iterations}
\label{sec:iter}
In order to extract the DNM gas present in both the dust and \g-ray data, we iteratively coupled the \g-ray and dust models. We started by fitting the dust optical depth and the \g-ray data independently according to Eqs. \ref{eq:modDust} and \ref{eq:modGam} with only the \hi maps as gaseous components and with ancillary (other than gas) components.
We then built maps of positive residuals (see how below) between the data (dust or \g rays) and the best-fit contributions of the model components. We separated these excesses into two DNM maps according to their location, one in the Reticulum cloud and the other one in the North Hydrus area. 

The next loops in the iterative process provided the DNM templates estimated from the dust to the \g-ray model ($\tau_{353}^{\rm{DNM}}$ in Eq. \ref{eq:modGam}) and conversely provided the DNM column density maps derived from the \g rays to the dust model ($N_{\rm H}^{\rm{DNM}}(l,b)$ in Eq. \ref{eq:modDust}). 
We repeated the iteration between the \g-ray and dust models until the log-likelihood of the fit to \g-ray data saturated.

The estimates of the $q$ and $y$ model coefficients and of the DNM templates improve at each iteration since there is less and less need for the other components, in particular the \hi ones, to compensate for the missing gas. They can still miss gas at some level because the DNM maps provided by the \g rays or dust emission have limitations (e.g. dust emissivity variations, limited \g-ray sensitivity).

Care must be taken in the extraction of the positive residuals because of the noise around zero. A simple cut at zero would induce a positive offset bias, and so we denoised the residual maps using the multi-resolution support method implemented in the MR filter software \citep{Starck98}. 
For the dust residuals, we used a multi-resolution thresholding filter, with six scales in the B-spline-wavelet transform (à trous algorithm), Gaussian noise and a 2-$\sigma$ threshold to filter. For the \g-ray residuals, we used the photon residual map obtained in the 0.16--63 GeV band and we filtered it with six wavelet scales, Poisson noise, and a 2-$\sigma$ threshold. 
The stability of the iterative analysis and the results of the dust and \g-ray fits are discussed in the following section.

\section{Results}\label{sec:results}
Before presenting the \g-ray emissivity spectra obtained in the Reticulum cloud in Sect. \ref{sec:CRret}, we discuss how the fits to the \g-ray and dust data improve with optical depth corrections to the \nhi column densities (Sect. \ref{sec:hicorr}) and with the addition of DNM gas templates (Sect. \ref{sec:DNM}). We also assess the robustness of the best-fit models with jackknife tests in Sect. \ref{sec:jack}.

\subsection{\hi optical depth correction}
\label{sec:hicorr}

We do not know the exact level of \hi opacity in the different clouds. This opacity induces systematic uncertainties in the \nhi column densities, and therefore also in the \hi contributions to the \g-ray and dust models.
\cite{Nguyen19} showed that a uniform isothermal correction of the \hi emission spectra across a cloud provides a good approximation to more precise estimates based on emission and absorption \hi spectra. 
We therefore produced the dust and \g-ray models for a wide range of optical depth corrections, from a rather optically thick case (\ts = 100 K) to the optically thin case that yields the minimum amount of gas in the cloud. 
Given the moderate \nhi column densities of the present clouds, the difference between those extreme corrections reaches 25\% on a few pixels and less than 1\% on the mean column density of a cloud.

\begin{figure}[!ht]
  \centering                
  \includegraphics[width=\hsize]{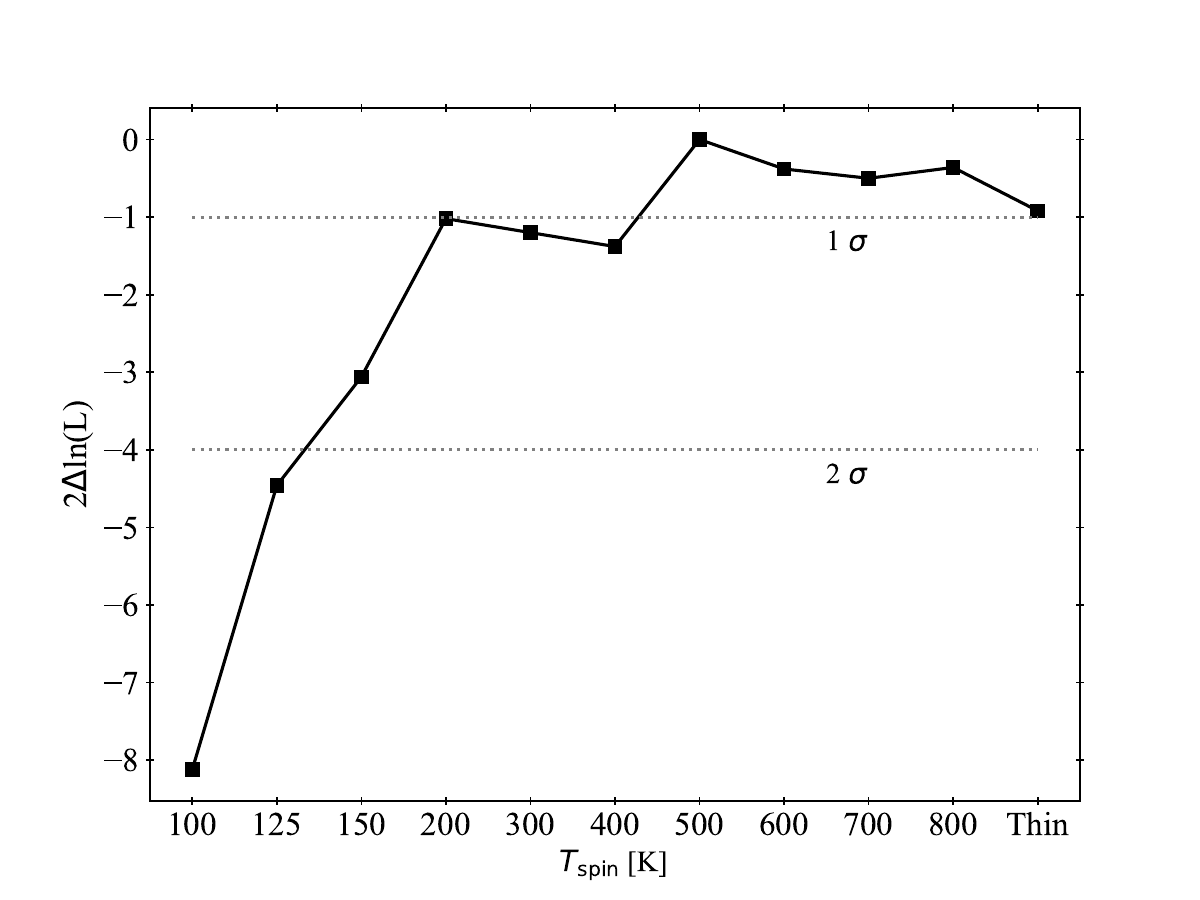}
  \caption{ Evolution of the log-likelihood ratio of the \g-ray fit in the 0.16--63~GeV band as a function of the \hi spin temperature used to calculate the \nhi column densities in the clouds.}
   \label{fig:dlnL_Tpsin}
\end{figure}

Even though the spin temperature is uniform, the \hi opacity correction depends on the \hi intensity recorded in each direction and velocity channel, so it varies across the cloud and slightly modifies the \nhi spatial distribution of a cloud. 
Figure \ref{fig:dlnL_Tpsin} shows that the goodness of the \g-ray fit (after the DNM iteration) significantly improves as we increase the \hi spin temperature. The \nhi maps best match the spatial structure of the \g-ray flux emerging from the atomic gas for \ts $\gtrsim 300$ K; therefore for small \hi opacities. Given the saturation of the log-likelihood ratio towards the optically thin case, we present the analysis results obtained for this thin case in the rest of the paper, unless otherwise mentioned. The residual maps between the \g-ray data and best-fit model show that using the same \ts for all the clouds can reproduce the \g-ray data without any significantly deviant cloud.

As the spin temperature \ts increases, the optical depth correction and \nhi column densities decrease, and the \g-ray emissivity $q_{\rm HI}$ and dust opacity $y_{\rm HI}$ of the \hi clouds increase. 
They do so respectively by 4--11\% and 3--7\% as the spin temperature rises over the entire range. \hi optical depth corrections have therefore a small effect in this region because the clouds are not massive, so they are fairly transparent to \hi line radiation.

\subsection{DNM detection}
\label{sec:DNM}

\begin{figure*}[!t]
  \centering   
  \includegraphics[width=\hsize]{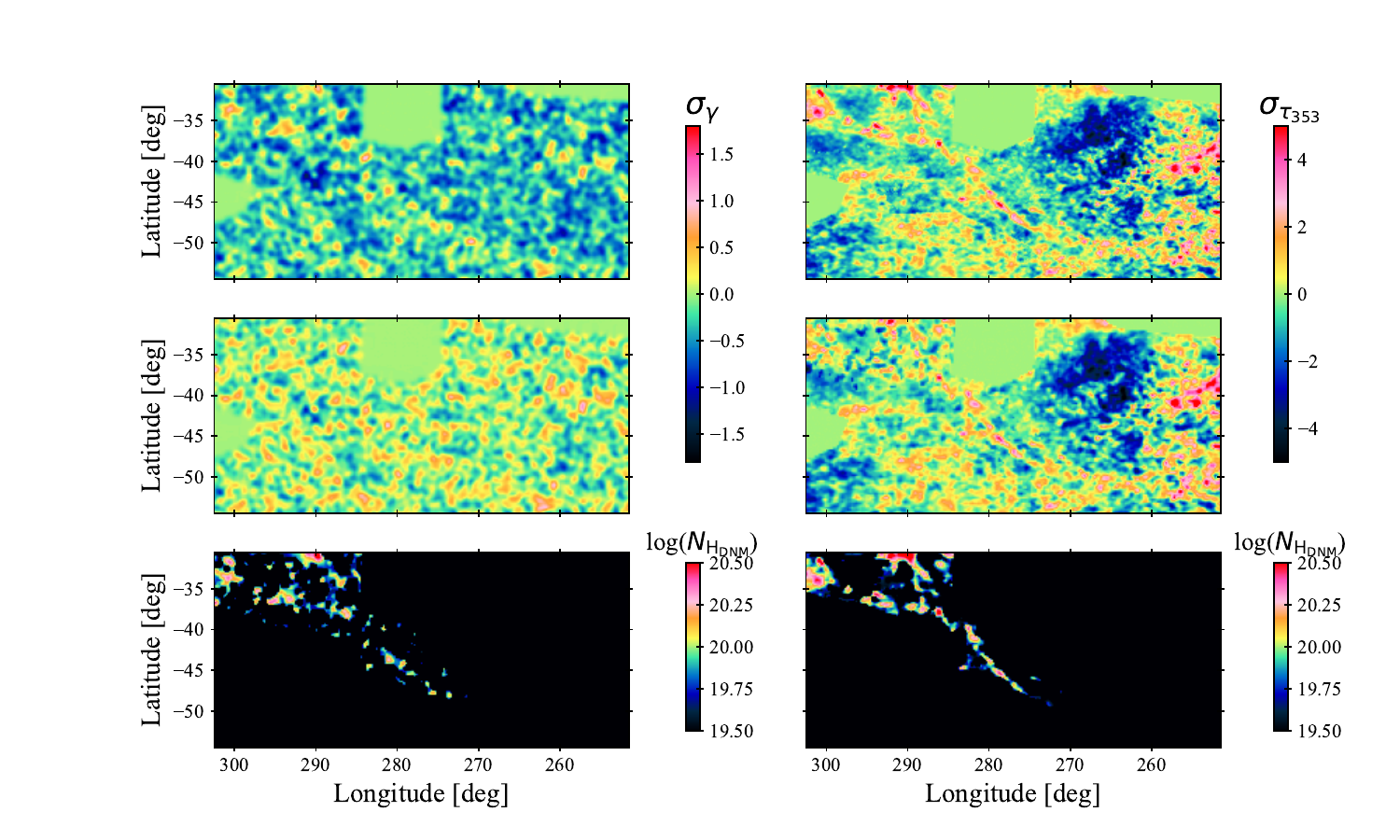}
  \caption{ Residual maps (data minus best-fit model,  in sigma units) in \g-ray counts in the 0.16--63~GeV band ($\sigma_{\gamma}$) and in dust optical depth ($\sigma_{\tau_{353}}$), obtained when including different sets of gaseous components in the models: without (\textit{top}) and with (\textit{middle}) the DNM components in addition to the \hi templates. The \nhdnm column-density maps (\textit{bottom}) extracted from the \g-ray residuals (\textit{left}) and from dust residuals (\textit{right}) are shown in the bottom row. The models are derived for an \hi at the optically thin case. }  
  \label{fig:allRes}
\end{figure*}

In order to check for the presence of DNM gas not traced by \hi line intensities, we first performed the dust and \g-ray fits with only the \nhi maps as gaseous components. The residuals are displayed in the upper panels of Fig. \ref{fig:allRes}. Positive residuals indicate that the data exceed the model predictions. Coincident excesses are seen in dust and in \g rays even though the two fits are independent. They cross the region in diagonal, from the North Hydrus cloud (upper left corner) down along the Reticulum filament. A coincident increase in CR density and in dust opacity is very improbable. Hadronic CR interactions with dust grains produce undetectable \g-ray emission \citep{Grenier05}, so this correlated excess is likely due to additional gas. The distribution of positive \g-ray excesses along the Reticulum cloud is sparse, but the large extent of negative (blue) excesses across the whole analysis region indicates that the \g-ray fit has unduly pushed the \hi emissivities of the clouds upward to compensate for some missing gas.

The DNM gas column density is built by correlating and filtering the dust and \g-ray excess signals in the iteration between both model fits (see Sect. \ref{sec:iter}). We built separate DNM maps for the North Hydrus and Reticulum clouds. Adding them improves the fits quality (see the middle maps of Fig. \ref{fig:allRes}). A log-likelihood ratio increase by 161 between the \g-ray fits with and without the DNM templates implies a very significant DNM detection. 
The spectrum of the \g-ray emission associated with both DNM components is consistent with that found in the \hi gas phase over the whole energy range (see below). This gives further support to a gaseous origin of the \g-ray emission in the DNM structures. 

Positive residuals remain in the final dust fit, even after the addition of the DNM maps. Those in spatial coincidence with the DNM phase are partially due to the non-linear increase in dust opacity as the gas density rises towards the dense molecular phase \citep{Remy17}. They also result from the limited \g-ray photon statistics that limit the accuracy of the DNM template provided to the dust fit. The negative and positive dust residuals seen at longitudes $l < 270^\circ$ reflect limitations in the \taunu data map due to the difficult separation of the dust emission from other submillimetre emission in these very faint zones \citep[][ and references therein]{Planck14XI,Irfan19}. These regions were excluded from the DNM analysis.

\subsection{Best fits and jackknife tests}
\label{sec:jack}

% *****************************************************
\begin{table*}
\caption{Best-fit dust opacities in the \hi and DNM phases of the Reticulum and North Hydrus clouds, and isotropic optical depth.}
\centering
\begin{tabular}{ccccc}
\hline
$y_{\rm{HI,Ret}}$ [10$^{-27}$~cm$^{2}$] & $y_{\rm{DNM,Ret}}$ [10$^{-27}$~cm$^{2}$] & $y_{\rm{HI,HyiN}}$ [10$^{-27}$~cm$^{2}$] & 
$y_{\rm{DNM,HyiN}}$ [10$^{-27}$~cm$^{2}$] & $y_{\rm iso}$ \\ 
\hline
8.91$\pm$0.07$\pm$0.12 & 5.05$\pm$0.40$\pm$0.58 & 8.68$\pm$0.03$\pm$0.05 & 14.90$\pm$0.42$\pm$0.44 &
(2.79$\pm$0.06$\pm$0.12) 10$^{-7}$ \\ 
\hline \\
\end{tabular}
\tablefoot{The first uncertainties are statistical (1$\sigma$) and the second give the standard deviations obtained in the jackknife fits. }
\label{tab:yfit}
\end{table*}
%[] & [10$^{-27}$~cm$^{2}$] & [10$^{-27}$~cm$^{2}$] & [10$^{-27}$~cm$^{2}$] & [10$^{-7}$] \\ % 
% *****************************************************
\begin{table*}
\caption{Best-fit \g-ray emissivities spectra per gas nucleon in the \hi and DNM phases of the Reticulum and North Hydrus clouds.}
\centering
\begin{tabular}{ccccc}
\hline
Energy [MeV] & $q_{\rm{HI,Ret}}$ & $q_{\rm{DNM,Ret}}$ [10$^{26}$~cm$^{-2}$] & $q_{\rm{HI,HyiN}}$ & $q_{\rm{DNM,HyiN}}$ [10$^{26}$~cm$^{-2}$] \\ 
\hline
10$^{2.2}$ -- 10$^{2.4}$ & 1.56$\pm$0.23$\pm$0.09 & 0.21$\pm$1.22$\pm$0.28 & 1.17$\pm$0.15$\pm$0.04 & 0.78$\pm$0.52$\pm$0.18 \\ 
10$^{2.4}$ -- 10$^{2.6}$ & 1.27$\pm$0.13$\pm$0.07 & 0.74$\pm$0.61$\pm$0.26 & 1.10$\pm$0.08$\pm$0.03 & 0.61$\pm$0.26$\pm$0.09 \\ 
10$^{2.6}$ -- 10$^{2.8}$ & 0.98$\pm$0.10$\pm$0.06 & 0.85$\pm$0.41$\pm$0.20 & 1.02$\pm$0.06$\pm$0.02 & 0.65$\pm$0.17$\pm$0.06 \\ 
10$^{2.8}$ -- 10$^{3.0}$ & 1.16$\pm$0.08$\pm$0.04 & 0.85$\pm$0.29$\pm$0.12 & 1.04$\pm$0.04$\pm$0.02 & 0.77$\pm$0.12$\pm$0.04 \\ 
10$^{3.0}$ -- 10$^{3.2}$ & 0.98$\pm$0.07$\pm$0.04 & 0.85$\pm$0.27$\pm$0.14 & 1.03$\pm$0.04$\pm$0.01 & 0.64$\pm$0.11$\pm$0.05 \\ 
10$^{3.2}$ -- 10$^{3.6}$ & 0.98$\pm$0.07$\pm$0.04 & 1.46$\pm$0.26$\pm$0.11 & 1.00$\pm$0.04$\pm$0.02 & 0.70$\pm$0.11$\pm$0.04 \\ 
10$^{3.6}$ -- 10$^{4.0}$ & 1.02$\pm$0.16$\pm$0.09 & 1.20$\pm$0.55$\pm$0.32 & 1.12$\pm$0.09$\pm$0.03 & 0.98$\pm$0.23$\pm$0.09 \\ 
10$^{4.0}$ -- 10$^{4.8}$ & 0.93$\pm$0.38$\pm$0.18 & 0.05$\pm$1.14$\pm$0.25 & 1.13$\pm$0.21$\pm$0.06 & 1.18$\pm$0.55$\pm$0.17 \\ 
10$^{2.2}$ -- 10$^{4.8}$ & 1.11$\pm$0.04$\pm$0.03 & 0.99$\pm$0.15$\pm$0.09 & 1.05$\pm$0.02$\pm$0.01 & 0.75$\pm$0.07$\pm$0.03 \\ 
\hline
\end{tabular}
\tablefoot{The first uncertainties are statistical (1$\sigma$) and the second give the standard deviations obtained in the jackknife fits. }
\label{tab:qfit}
\end{table*}
%[MeV] & & [10$^{26}$~cm$^{-2}$] & & [10$^{26}$~cm$^{-2}$] \\ % *****************************************************

\begin{figure*}%[!ht]
  \centering                
  \includegraphics[width=\hsize]{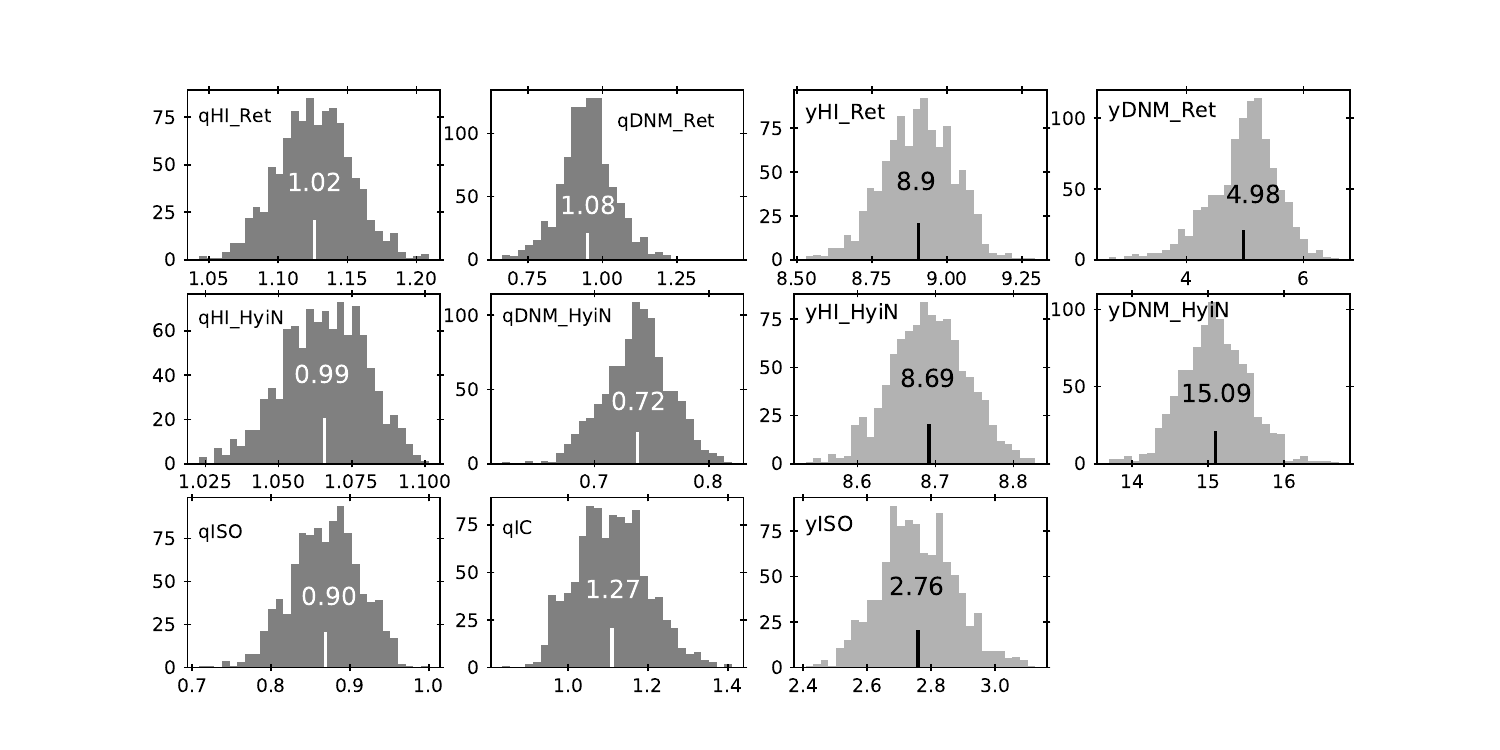}
  \caption{ Distributions of the best-fit coefficients obtained in the thousand jackknife fits to the 0.16--63~GeV \g-ray data (two left columns) and dust data (two right columns) for the optically thin \hi case. The vertical lines and overlaid numbers give the mean values of the distributions. The units are $10^{26}$ cm$^{-2}$ for $q_{\rm DNM}$, $10^{-27}$ cm$^{2}$ for $y_{\rm HI}$ and $y_{\rm DNM}$, and $10^{-7}$ for $y_{\rm iso}$. $q_{\rm{HI,i}}$, $q_{\rm{IC}}$ and $q_{\rm iso}$ are dimensionless normalisation factors. 
  }  
  \label{fig:qyjack}
\end{figure*}

The best dust and \g-ray fits with the models described by Eqs. \ref{eq:modDust} and \ref{eq:modGam} include the nine \hi and two DNM maps for the gaseous components. Tables \ref{tab:yfit} and \ref{tab:qfit} give the best-fit coefficients obtained for the \hi and DNM phases of the main Reticulum and North Hydrus clouds. The results found for the other small local clouds will be discussed in a subsequent paper (their emissivity spectra are rather close to the \qlis expectation). We focus here on the Reticulum filament to compare it with the Eridu one and we include the emissivity measurement in the North Hydrus cloud as it partially overlaps the northern end of Reticulum.
The \g-ray emissivities are given in each of the eight energy bands and for the whole 0.16--63~GeV band. 

The final iteration of the dust and \g-ray fits leads to the residual maps presented in the middle panels of Fig. \ref{fig:allRes}. 
The \g-ray residuals are small and randomly distributed, so our linear model provides an excellent statistical fit to the \g-ray data in the 0.16--63~GeV energy band across the whole region. The same conclusion is reached in the eight separate energy bands that are not shown here. 

The statistical errors on the best-fit coefficients are inferred from the information matrix of the fits \citep{Strong85}. They include the correlation between parameters. They are generally small (2--9\% for the \hi gas \g-ray emissivities, 15\% and 10\% for DNM emissivities for Ret and HyiN, 13\% for $q_{\rm IC}$, 9\% for $q_{\rm iso}$, 0.4--1.3\% for the \hi dust opacities, 8\% and 3\% for $y_{\rm DNM,Ret}$ and $y_{\rm DNM,HyiN}$ and 2\% for $y_{\rm iso}$) because the clouds are spatially well resolved in the analysis and because tight correlations exist between the gas, dust and \g-ray maps (see Fig. \ref{fig:gamDust}). 

We also checked the magnitude of potential systematic uncertainties due to the linear-combination approximation underlying the models. In order to evaluate the impact of possible spatial changes in the model parameters and/or in the mean level of \hi self- absorption, we repeated the last iteration of the dust and \g-ray fits a thousand times over random subsets of the analysis region, each time masking out 20\% of the pixels with 1$^\circ$-wide randomly selected squares. In \g rays, the jackknife tests were performed for the total 0.16–-63~GeV energy band and for each of the eight bands.
Figure \ref{fig:qyjack} shows the distribution obtained for the best-fit coefficients relating to the Reticulum and North Hydrus clouds and to the large-scale IC and isotropic components. The Gaussian-like distributions indicate that the emissivities and opacities presented in Tables \ref{tab:yfit} and \ref{tab:qfit} are statistically stable and that they are not driven by subset regions in each cloud complex. The standard deviations found in the jackknife distributions amount to 1--6\% for the \hi gas \g-ray emissivities, 9\% and 4\% for the Ret and HyiN DNM emissivities, 8\% for $q_{\rm IC}$, 5\% for $q_{\rm iso}$, 0.6--3\% for the \hi dust opacities, 12\% and 3\% for $y_{\rm DNM,Ret}$ and $y_{\rm DNM,HyiN}$ and 4\% for $y_{\rm iso}$. 
Both statistical errors (1$\sigma$) and systematic uncertainties (jackknife standard deviations) are given in Tables \ref{tab:yfit} and \ref{tab:qfit}. 
We quadratically added them to construct the final uncertainties on the $q$ and $y$ coefficients. 

\begin{figure*}[!ht]
  \centering     
  \includegraphics[scale=0.8]{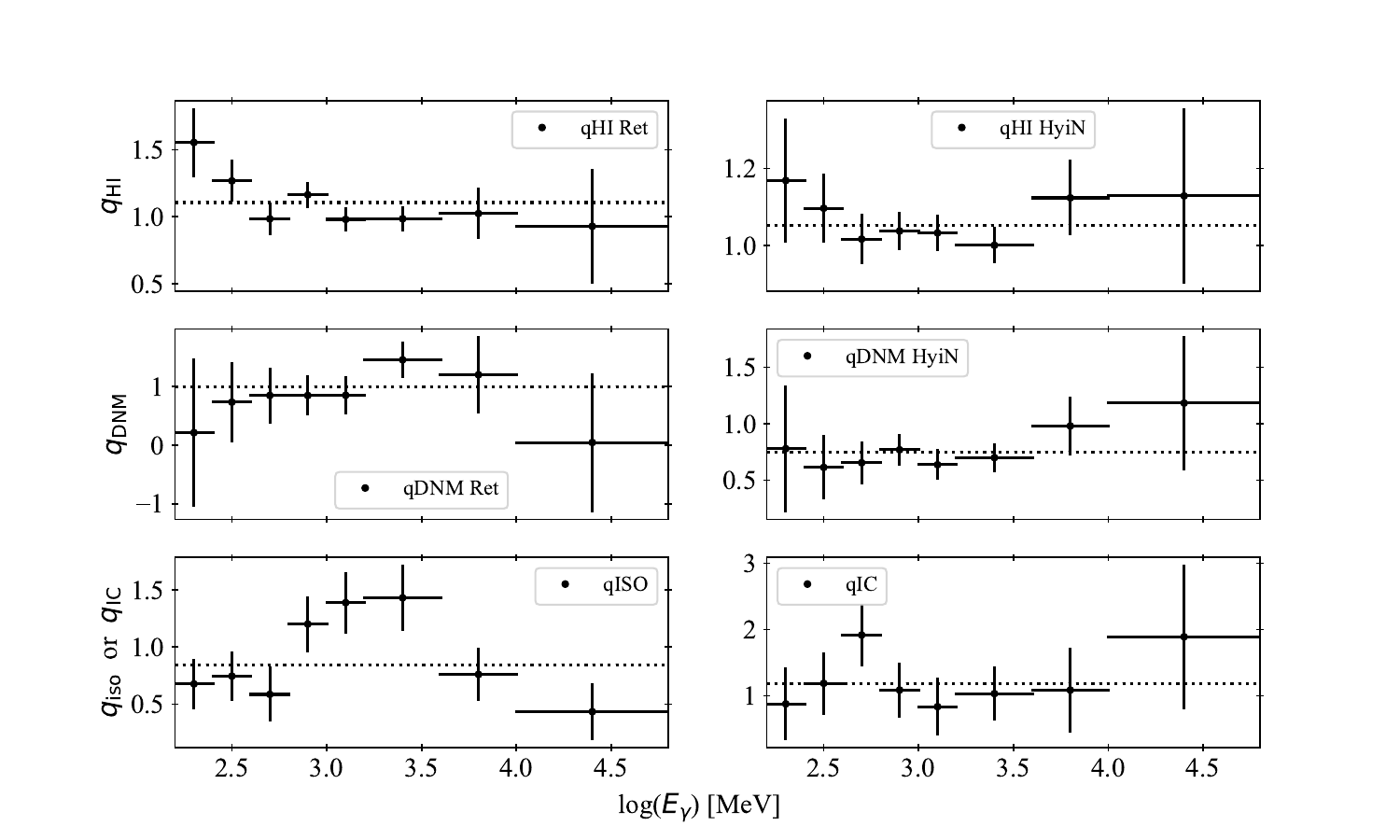}
  \caption{ Spectral evolution relative to the local interstellar spectrum \qlis of the \g-ray emissivities found for the \hi and DNM phases of the Reticulum and North Hydrus clouds (top and middle panels). Spectral evolution of the isotropic and IC intensities found in the region relative to the input spectra (bottom panel). $q_{\rm DNM}$/\qlis is given in $10^{26}$~cm$^{-2}$. The results are shown for the optically thin \hi case. The error bars combine the statistical and systematic uncertainties. } 
  \label{fig:qHIqDNM_energy}
\end{figure*}

Figure \ref{fig:qHIqDNM_energy} shows the stability with \g-ray energy of the best-fit coefficients relating to the Reticulum and North Hydrus clouds and to the IC and isotropic intensities. We find no significant deviation with respect to the average coefficient fitted for the entire energy band (dashed lines). This implies that the \hi and DNM gas emissivity spectra in the two clouds have the same shape as the average $q_{\rm LIS}(E)$ spectrum measured in the local ISM. We further discuss their relative normalisation with respect to \qlis in the next section. 

Figure \ref{fig:qHIqDNM_energy} and the value of the relative intensity ratio $q_{\rm IC}=1.18\pm 0.15$ in the whole energy band further show that the IC spectrum detected in these directions of the sky is fully consistent in shape and in intensity with the GALPROP prediction based on the complex evolution of the interstellar radiation field and of the CR lepton spectra across the Milky Way. The isotropic spectrum found in this small region of the sky is also close to the larger-scale estimate obtained between 10$^\circ$ and 60$^\circ$ in latitude\footnote{http://fermi.gsfc.nasa.gov/ssc/data/access/lat/BackgroundModels.html}.  
%emissivity 
The relative intensity ratio is $q_{\rm iso}=0.84\pm 0.08$ in the 0.16--63~GeV energy band. The interstellar foregrounds are faint enough to separate the IC and isotropic components despite rather similar spatial distributions across the analysis region (see Fig. \ref{fig:NHI}).

\cite{Planck14XI} and \cite{Remy17} found marked gradients in dust opacity across the sky. These variations reflect a systematic opacity increase as the gas becomes denser and the grains evolve in size and chemistry \citep[][and references therein]{Remy17,Ysard19}. Diffuse atomic cirri at high latitudes exhibit low opacities around a mean value of $(7.0 \pm 2.0) \times 10^{-27}$ cm$^2$ \citep{Planck14XI} or $(7.1 \pm 0.6) \times 10^{-27}$ cm$^2$ \citep{Planck14XVII}) whereas larger values ranging from 10 $\times 10^{-27}$ to 16 $\times 10^{-27}$ cm$^2$ are found in the denser atomic envelopes of nearby clouds \citep{Remy17}, and values from 15 $\times 10^{-27}$ to 17 $\times 10^{-27}$ cm$^2$ characterise the DNM phase \citep{Planck15_Cham,Remy17,Joubaud20}. Dust opacities further increase in the dense, CO-bright, molecular phase, reaching values as high as $60 \times 10^{-27}$ cm$^2$ \citep{Remy17}. The dust opacities we found in the \hi phase of the Reticulum and North Hydrus clouds are consistent with this trend. They are close to the low diffuse cirrus values and, in particular, to the dust opacity of $(6.4 \pm 0.1) \times 10^{-27}$ cm$^2$ measured in the Eridu filament \citep{Joubaud20}. The dust opacity we found in the DNM phase of the North Hydrus cloud fully agrees with other estimates. We caution the reader that the unusually low $y_{\rm DNM}$ value found for the Reticulum cloud may not be reliable because of the difficulty discussed in Sect. \ref{sec:DNM} to fully extract the \nhdnm column density along Reticulum from the \g-ray data. The final dust residual map in Fig. \ref{fig:allRes} shows a leftover positive excess along the Reticulum filament. Taking this excess into account would increase the $y_{\rm DNM}$ value. 

\subsection{Cosmic rays in the Reticulum and Eridu filaments}
\label{sec:CRret}

\begin{figure}[!t]
  \centering     
  \includegraphics[scale=0.48]{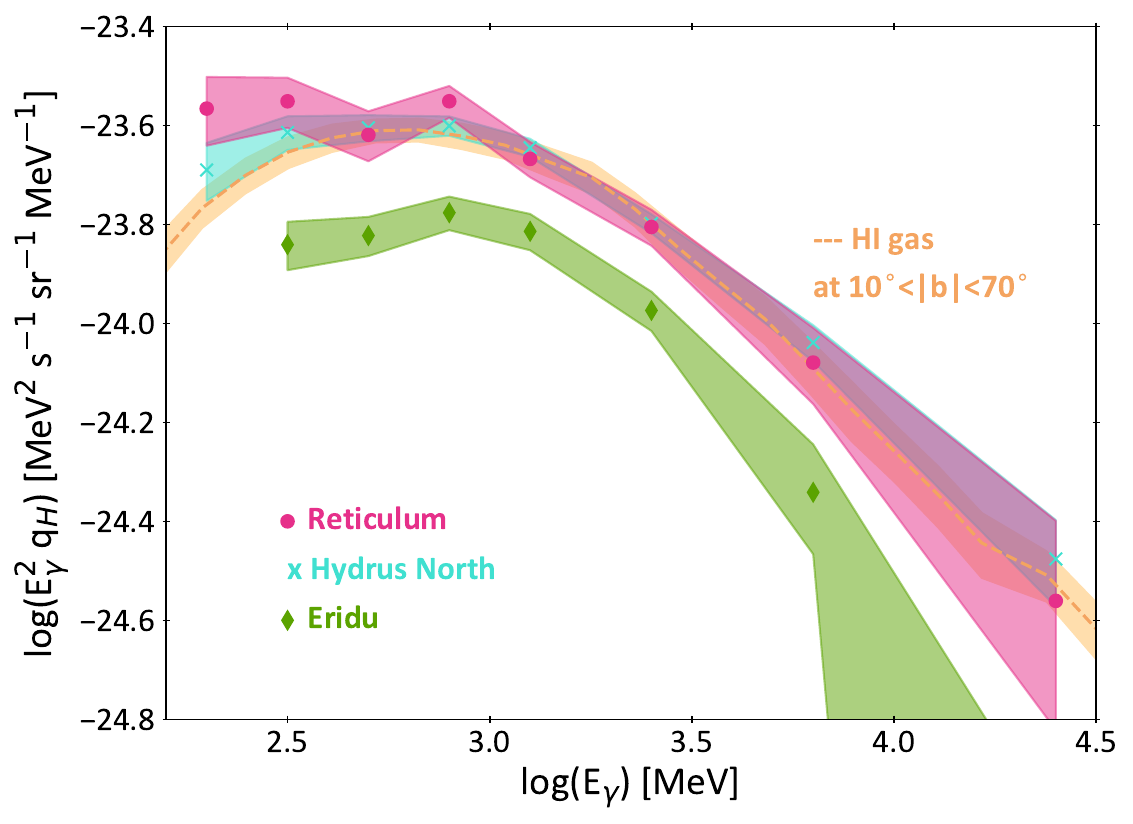}
  \caption{ Spectral evolution of the \g-ray emissivities per gas nucleon in the Eridu (green), Reticulum (red), and North Hydrus (cyan) clouds. The orange curve gives the average \qlis emissivity measured in the atomic gas at latitudes $10^{\circ} \leq |b| \leq 70^{\circ}$ \citep{Casandjian15}. The emissivities are shown for the \hi optical-depth correction that best fits the \g-ray data (\ts = 100~K for Eridu, 140~K for \qlis, and optically thin for the others). The shaded bands include the statistical and systematic uncertainties. }   
  \label{fig:spectreRetEri}
\end{figure}

Figure \ref{fig:spectreRetEri} compares the \g-ray emissivity spectrum we found per gas atom in the Reticulum filament to that in the similar filament of Eridu and in the more massive North Hydrus cloud. We also compare it to the average \qlis spectrum measured in the local atomic gas at latitudes between 10$^{\circ}$ and 70$^{\circ}$ across the sky \citep{Casandjian15}. We compare them for the spin temperatures that best fitted their respective \g-ray data; therefore at \ts = 100~K for Eridu, 140~K for \qlis, and for optically thin \hi for Reticulum and North Hydrus. The results indicate that the CR population that permeates the \hi and DNM phases of the Reticulum and North Hydrus clouds has the same energy distribution and same flux as the average population in the local ISM. This population applies to the 100-150~pc long sector subtended by the Reticulum and North Hydrus clouds along the rim of the Local Valley. As expected for the low \nhi column densities characterising these clouds, we find no significant spectral deviation at low energies that would mark a loss or concentration of CRs when penetrating into the denser DNM clumps \citep{Schlickeiser16,Bustard21}.

Figure \ref{fig:spectreRetEri} illustrates that the CR flux in the Reticulum filament is $1.57 \pm 0.09$ times that pervading the similar Eridu filament. This discrepancy decreases to $1.51 \pm 0.09$ if we compare them for the same \hi spin temperature of 100~K, to $1.38 \pm 0.08$ if we assume them both optically thin. The minimum ratio is $1.32 \pm 0.08$ if Eridu is optically thin and Reticulum thicker with \ts = 100~K, a case yielding a worse \g-ray fit to the data for both clouds.
The loss mechanism that affects CRs in Eridu should be energy independent to preserve the same emissivity spectral shape in the three clouds across the entire energy band. 

\halpha hydrogen recombination line emission has been recorded along the Reticulum cloud in the Southern Halpha Sky Survey Atlas \citep{Gaustad01}, suggesting that the cloud be a cooling filament or cooling shell surrounded by recombining warm ionised gas. The spatial correlation between the dust and \g rays detects any gas in addition to the \hi gas, regardless of its chemical or ionisation state. The inferred DNM column densities are dominated by the dense CNM and diffuse \hd, but they also include the much more dilute ionised gas. The \g-ray emissivity we obtained in the \hi phase of Reticulum is therefore not strongly biased by untraced ionised gas surrounding it. As a further check, we used the 
%6\arcmin~(FWHM) resolution 
\halpha map reprocessed by \cite{Finkbeiner03} as an additional gas template into the dust and \g-ray fits, even though the \halpha emission measure does not scale with the \nhp column density, but it integrates the product of the electron and ion densities along the line of sight,
$\text{EM}_\alpha \propto \int_{LOS} n_{\rm e} n_{\rm H^+} dl$. With this crude additional template, the Reticulum \g-ray emissivity remains consistent with the local-ISM \qlis average and is still $1.51\pm 0.11$ times larger than in Eridu over the whole energy band for the choice of best fitting spin temperatures.  

Assuming that the same CR flux pervades the recombining and neutral hydrogen, the \g-ray signal related to the \halpha template yields \nhp column densities around $10^{19}$~cm$^{-2}$ in Reticulum, an order of magnitude below those in the neutral phase. Column densities of the Galactic warm ionised medium, integrated to the Reticulum distance, are also an order of magnitude below the \nhi ones and they would not spatially correlate with the \hi clouds to impact the values of the $q_{\rm HI}$ emissivities \citep{Gaensler08}. 

\section{Discussion}

The Reticulum and Eridu clouds are both light and filamentary neutral clouds, with ordered magnetic fields on a few parsec scale, and a large inclination with respect to the Galactic plane. Yet, the first has 57 $\pm$ 9\% more CRs than the second. In the next sections, we try to characterise the dynamical and magnetic states of those clouds at a parsec scale in order to search for differences that can influence CR transport inside them at the much smaller scales of their gyro-motion. CR momenta parallel to the local magnetic field randomly vary by interaction with magnetic perturbations. The Alfvén and slow modes of interstellar magnetohydrodynamic (MHD) turbulence are thought to be inefficient scatterers because of their large anisotropy at small scales \citep{Chandran00,Xu16}. We therefore first focus on gas properties that can impact CR diffusion in the self-confinement scenario where the particles scatter on Alfvén waves they generate by the streaming instability \citep{Kulsrud69}. Both interstellar and self-excited MHD turbulence should be severely damped by ion--neutral interactions in these neutral clouds, but CR gyration along tangled magnetic-field lines as well as the perpendicular (super)diffusive wandering of the lines also induce an effective CR diffusion in space, with a mean free path close to the magnetic-field coherence length \citep{Lazarian16,Hu22,Sampson23}. We therefore compare the dispersion in gas velocity and in magnetic bending in the two clouds. All CR transport modes strongly depend of the state of the MHD turbulence, so we derive the \bsky magnetic-field strength from dust polarisation observations and we compare the sonic Mach numbers and gas-to-magnetic pressure ratios along the cores of both filaments.

\subsection{Gas densities in the atomic phase}
\label{sec:densities}

\begin{figure}[!ht]
  \centering     
  \includegraphics[scale=0.75]{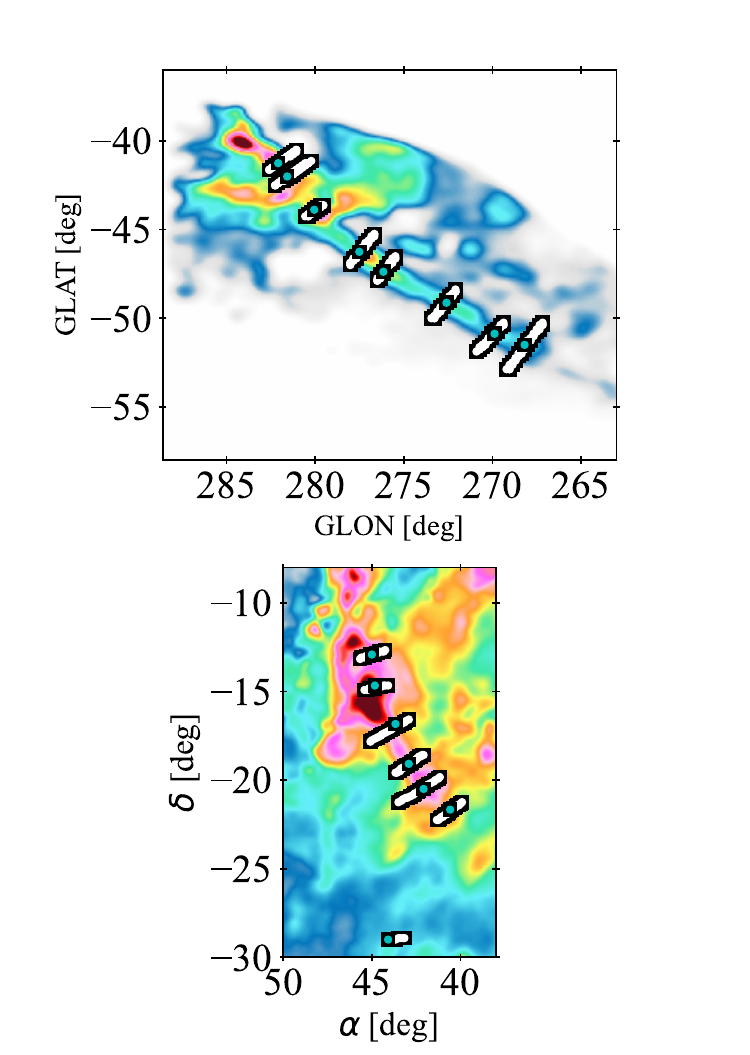}
  \caption{\nhi maps of the Reticulum (top) and Eridu (bottom) filaments, smoothed by a Gaussian kernel of 0.3$^\circ$ and 0.2$^\circ$, respectively, to highlight the cloud edges. The white lines show the perpendicular cuts used to estimate the cloud widths. The cyan squares show $1^\circ$-wide squares adopted to estimate gas volume densities, \bsky field strengths and parallel diffusion coefficients $\kappa_\parallel$ along the filament core. }  
  \label{fig:CloudWidth}
\end{figure}

\begin{figure}[!ht]
  \centering     
  \includegraphics[scale=0.6]{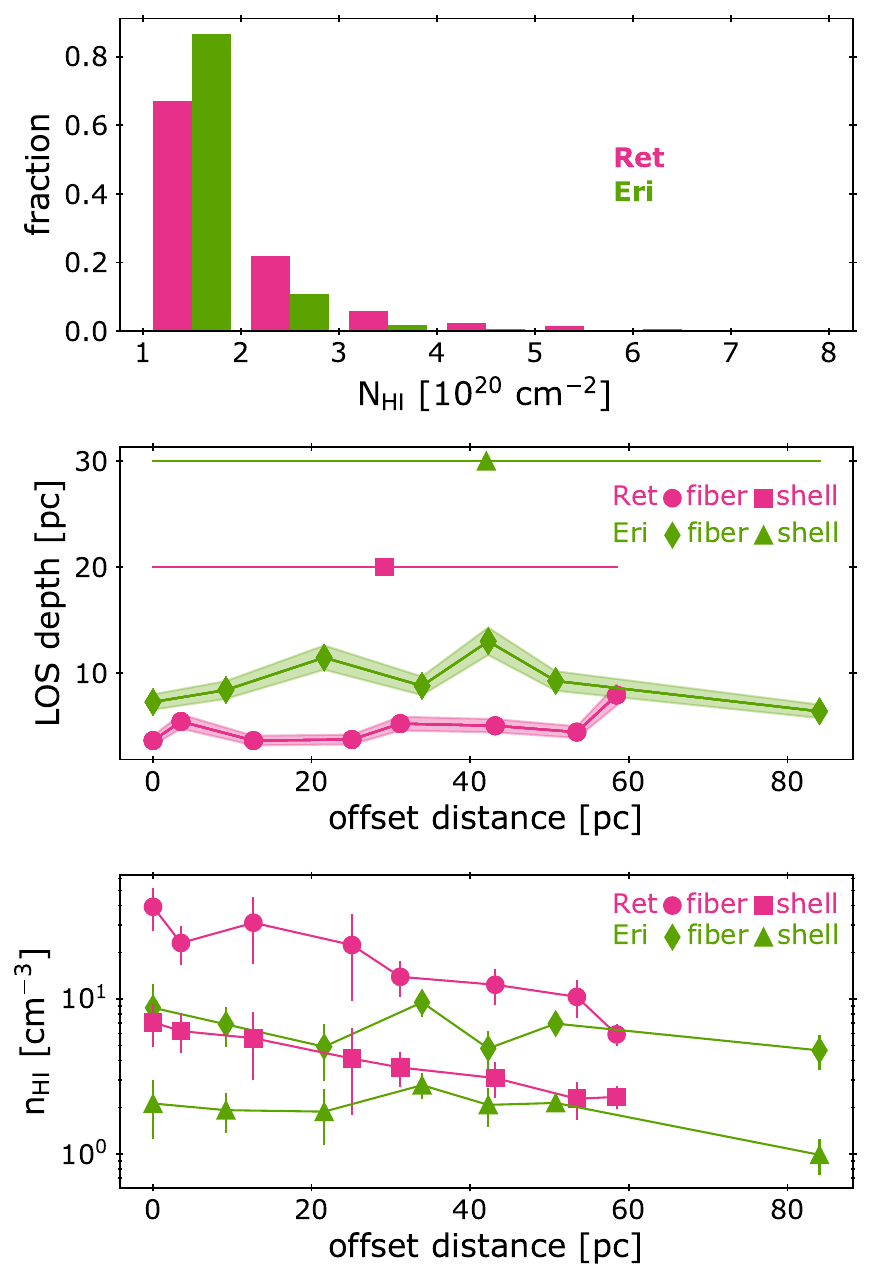}
  \caption{Distributions of \hi column densities (top), LOS depths (middle), and mean \hi volume densities (bottom) in the Reticulum (magenta) and Eridu (green) clouds. The offset distance is the separation between a $1^\circ$-wide square along a filament and the first one (of largest latitude or declination). The different symbols refer to the fibre or shell cloud geometry. The latter provides upper/lower limits to the depth/density. The LOS depth shading is due to the distance uncertainty. The data points and error bars in volume density respectively give the mean value in each ${\sim}5$~pc-wide square and the rms dispersion at 0.5--0.7~pc scale inside each square. }
  \label{fig:NHI_Ep_nH}
\end{figure}

Given the elongated aspect of the Reticulum and Eridu filaments, we used two different assumptions for the cloud geometry, namely a cylindrical fibre or a shell seen edge on. These geometries bracket the unknown depth along the line-of-sight (LOS) in order to estimate \hi gas volume densities, $n_{\rm HI}$. In the fibre configuration, the LOS depth is close to the observed width in the plane of the sky. In order to estimate the latter, we used the Gaussian full widths at half max (FWHM) of the \nhi column-density profiles observed along cuts perpendicular to the filament axis. Figure \ref{fig:CloudWidth} shows the eight and seven cuts respectively chosen along the Reticulum and Eridu filaments.  
The resulting depths are shown in the middle panel of Fig. \ref{fig:NHI_Ep_nH} with an uncertainty reflecting that in cloud distance. The mean depth and rms dispersion are $5.0\pm 0.6$~pc in Reticulum and $10.5\pm 1.0$~pc in Eridu. 
In the shell shaped configuration, the LOS depth is closer to half the chord length through the thick shell. Given the mean widths measured above and a shell radius of curvature of about 100 pc, we estimated the LOS depths to be roughly 20~pc for Reticulum and 30~pc for Eridu. These limits are represented as triangles in the middle plot of Fig. \ref{fig:NHI_Ep_nH}. They provide reasonable upper limits to the cloud depth.

We derived the mean gas volume density in a $1^\circ$-wide square centred on the peak of each \nhi profile cutting perpendicularly through the Reticulum and Eridu filaments. The results are shown in the bottom plot of Fig. \ref{fig:NHI_Ep_nH}. The shell geometry provides lower limits to the gas densities.

Figure \ref{fig:NHI_Ep_nH} illustrates that the larger \nhi column densities recorded in the Eridu filament are likely due to longer LOS depths and that the Reticulum filament has denser substructures than Eridu. In both geometries, the density estimates along Reticulum are typical of the CNM and they are consistent with the detection of comparable \nhdnm column densities in the DNM. For both geometries, the Eridu volume densities are more typical of the unstable lukewarm neutral medium (LNM) that transitions between the stable WNM and CNM phases. They indicate that CNM clumps are likely present inside both clouds, but should be more numerous and more densely packed in Reticulum than in Eridu. 

\subsection{Self-confined diffusion}
\label{sec:kappa}

\begin{figure*}[ht]
  \centering     
  \includegraphics[scale=0.7]{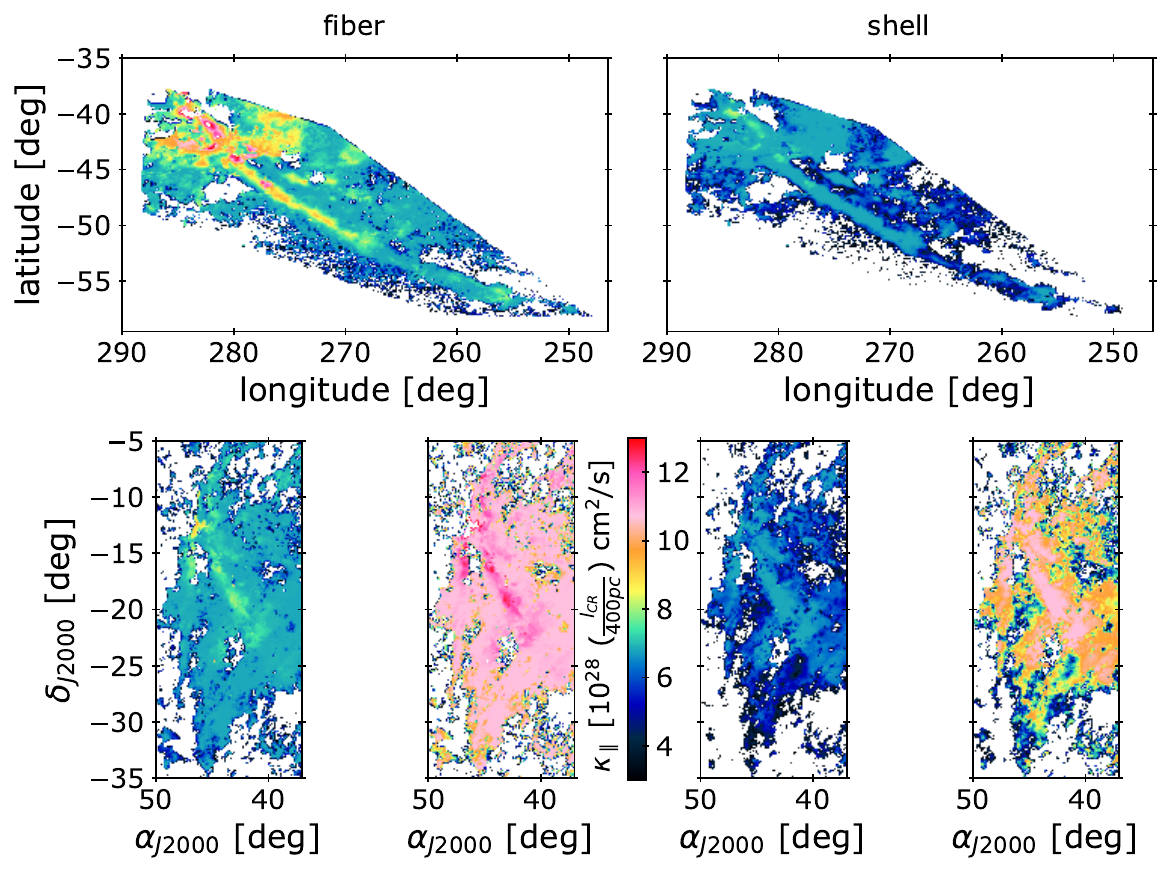}
  \caption{Parallel diffusion coefficients obtained for the Reticulum (top) and Eridu (bottom) clouds in the fibre (left) and shell (right) geometries, assuming the damping rate of Eq. \ref{eq:Soler}, a characteristic CR flux variation length $\ell_{\rm CR}= 400$ pc, and a uniform LOS depth equal to the shell one or to the average of the fibre ones. For each geometry, the Eridu plots show $\kappa_\parallel$ values for the same CR flux exciting the Alfvén waves as in Reticulum (left) or for the reduced CR flux corresponding to the cloud lower \g-ray emissivity (right).} 
  \label{fig:kappaMap}
\end{figure*}

CRs gyrate and flow along magnetic field lines. They scatter off magnetic fluctuations with sizes close to their gyro-radius (e.g. ${\sim}2 \,\mu$pc for a 10 GeV CR in a $B=5 \,\mu$G field). For instance, in the frame moving with Alfvén waves along the field, the Lorentz force acting on a CR during a gyro-resonant interaction changes the particle's pitch angle ($\tan \theta = \mathbf{p} \wedge \mathbf{B} / \mathbf{p} \cdot \mathbf{B}$), but not its energy, so random phase shifts between the gyro-motion and the encountered waves cause a random walk in pitch angle and reduce the CR effective speed along the field line. See \citep{Ruszkowski23} for a nice pictural description of gyro-resonant scatterings in pitch angle. Whereas very high energy CRs are likely scattered by interstellar MHD turbulence, the lower-energy CRs that we sample here in \g rays are more likely scattered by Alfvén waves that they self-generate via the resonant streaming instability \citep{Kulsrud69}. The anisotropy of interstellar MHD turbulence  increases with decreasing scale in the cascade \citep{Goldreich95}. The eddies become elongated along the field because of turbulent reconnection. As a result, during the resonant encounter of a CR with a gyro-radius-long eddy, the gyro-orbit perpendicular to the field encompasses other uncorrelated eddies that cancel out the net change in pitch angle \citep{Chandran00}. Unlike Alfvén and slow modes, fast compressive turbulent modes remain much more isotropic, and are therefore efficient scatterers on small scales. In the neutral atomic gas, however, they mostly affect CRs of much larger energies ($\gg 100$ TeV) than those of interest here \citep{Xu16}.  

In the self-confinement scenario, a slight anisotropy in the CR pitch-angle distribution can excite Alfvén waves with the right geometry to induce further pitch-angle scattering \citep{Kulsrud69}. The required level of anisotropy to trigger the instability corresponds to a bulk drift velocity $v_D$ larger than the ion Alfvén speed $v^{\rm ion}_A = B / \sqrt{4\pi \rho_i}$ (the drift velocity is that of the frame where the CR distribution is isotropic).   
The wave growth rate by this streaming instability depends on the integrated number density of exciting CRs and on the mass density of interstellar gas ions, $\rho_i$, as:  
\begin{equation}
    \Gamma_{\rm grow}(k) = \frac{\pi^{3/2}}{2} \frac{e}{c} (v_D-v^{\rm ion}_A) \frac{1}{\sqrt{\rho_i}} \sum_j \frac{\alpha_j - 3}{\alpha_j -2} Z_j \, n_j(\mathcal{R}_j > \frac{B}{k}) ,
    \label{eq:growth}
\end{equation}
where $B$ is the mean field strength and we add the contributions of different CR nuclei with charge $Z_j e$, rigidity $\mathcal{R}_j$, and volume number density spectrum $n_j(\mathcal{R})$ with a power-law spectral index $\alpha_j$ in the range of interest \citep{Kulsrud69,Kulsrud71}. The CR density is integrated in rigidity above the threshold corresponding to the resonance condition $\mathcal{R}_j = B/k$ for the wave number $k$. We considered the spectra of CRs with $Z_j \leq 28$ that reach the heliopause \citep{Boschini20}. They correspond to the \qlis emissivity discussed above. We integrated the spectra over 1 GV in rigidity and fitted the $\alpha_j$ spectral indices in the 1--20 GV range that contributes most to the growth rate.

The actual level of CR anisotropy that causes the streaming instability and the growth of scattering waves is extremely difficult to estimate from observations and from theory. We therefore follow \cite{Jiang18} and \cite{Bustard21} in expressing the net drift velocity $v_D-v^{\rm ion}_A$ with respect to the Alfvén wave frame as a diffusive flux: 
\begin{equation}
v_D-v^{ion}_A = \frac{\kappa_\parallel}{\ell_{\rm CR}} , 
\label{eq:drift}
\end{equation}
with the characteristic length for a change in CR flux, $\ell_{\rm CR}$, and the diffusion coefficient $\kappa_{||}$ parallel to the mean $B$ field.

Wave amplitudes and scattering rates are reduced by wave damping processes that depend on the properties of the background MHD turbulence and on ambient plasma conditions.
The main damping mechanism in the neutral gas phases that we have probed in \g rays is the ion--neutral one. Following \cite{Hopkins21gam} and \cite{Xu22}, we checked that the linear Landau, nonlinear Landau, and turbulent damping rates are two to three orders of magnitude smaller than the ion--neutral one in the typical environments of the Reticulum and Eridu clouds.

Momentum exchange between the ions and hydrogen or helium neutral atoms, all with the same equilibrium temperature $T$, leads to the damping rate given by \cite{Soler16} :
\begin{equation}
\Gamma_{\rm in,S16} = \sqrt{\frac{32}{9\pi} \frac{k_B T}{A_i m_{\rm H}}} n_{\rm HI} \left[ \sigma_{\rm iH} \sqrt{\frac{1}{1+A_i}} + \sigma_{\rm iHe} \frac{x_{\rm He}}{x_{\rm HI}} \sqrt{\frac{A_{\rm He}}{A_i+A_{\rm He}}} \right]
\label{eq:Soler}
\end{equation}
where $k_B$ is the Boltzmann constant, $m_{\rm H}$ is the H atom mass,  $x_{\rm HI}$ and $x_{\rm He}$ are the neutral hydrogen and helium abundance by number, relative to the total number of hydrogen nucleons, and $A_{\rm He}$=4 is the atomic weight of helium. The ion--neutral momentum-transfer cross sections are $\sigma_{\rm i H} = 10^{-14}$~cm$^2$ and $\sigma_{\rm i He} = 3 \times 10^{-15}$~cm$^2$ from \cite{Vranjes13}.

The ion abundance, $x_i$, and mean ion atomic weight, $A_i$, vary with the gas state \citep{Draine11}. The predominant form of ions gradually varies from H$^+$ in the WNM to about half H$^+$ and half C$^+$ in the CNM, and their total abundance decreases from about 2\% in the WNM down by two orders of magnitude in the CNM. We used the heating, cooling, chemistry, and ionisation model of \cite{Kim23} to calculate the equilibrium gas temperature $T$, ion abundance $x_i$, and average ion weight $A_i$ as a function of the gas densities observed in the neutral phase ($n_{\rm HI}$) of the two clouds, and for the interstellar conditions of the local ISM ($x_{\rm He} = 0.1$, $x_{\rm C} = 1.6 \times 10^{-4}$ in the gas phase, solar metallicity, primary CR ionisation rate of $2 \times 10^{-16}$ s$^{-1}$, and local FUV intensity). This allowed us to estimate the damping rates of Alfvén waves in different locations along both clouds. 

Another formalism for ion--neutral damping is given by \cite{Kulsrud69} :
\begin{equation}
\label{eq:K69}
    \Gamma_{\rm in,K69} = \frac{1}{2}\frac{n_nm_n}{m_n+m_i}\langle\sigma v \rangle_{in} ,
\end{equation}
where $n_{\rm n}$ is the neutral number density, $m_{\rm n}$ is the mean mass of neutrals, and $m_i=A_im_{\rm H}$ is the mean mass of ions. The ion--neutral momentum transfer rates are $\langle\sigma v \rangle_{\rm in}=3.25 \times 10^{-9}$~cm$^3$ s$^{-1}$ for collisions between H and H$^+$, $2.39 \times 10^{-9}$~cm$^3$ s$^{-1}$ between H and C$^+$, and $1.42 \times 10^{-9}$~cm$^3$ s$^{-1}$ between He and H$^+$ (from Table 2.1 of \cite{Draine11}). This formulation does not depend on the gas temperature. 

We adopted the \cite{Soler16} damping rate as the baseline for this work. Assuming a uniform LOS depth equal to the mean value of the estimates shown in Fig. \ref{fig:NHI_Ep_nH} in each geometry, we find rather uniform damping rates close to $6 \times 10^{-9}$~s$^{-1}$ across both clouds, in both geometries. The two damping rate formulations yield comparable values across the clouds in the shell geometry, but not in the denser fiber configuration where the \cite{Kulsrud69} formulation yields a few times larger rates than the \cite{Soler16} ones (one to three times larger along Eridu and one to five times larger along Reticulum). This difference is due to the drop in temperature and increase in ion atomic weight in the denser fibre gas in the \cite{Soler16} formulation. 

We further note that these rates likely over-predict the actual damping rate in the partially ionised plasma because they do not take the charged dust grains into account even though each grain carries a charge that varies from 0.1e (small grains) to 20e or 30e (large grains) in the atomic gas phases of interest here \citep{Draine04}. The presence of dust charges reduces the dissipation of Alfvén waves by at least an order of magnitude \citep{Hennebelle23}. Charged grains also lower the drift velocity required for CR excitation of the waves. 
The gas-to-dust mass ratio being of order 100, the ion Alfvén speed is limited to about ten times the gas one ($v_A^{ion}/v_A^{gas} \approx \sqrt{\rho/\rho_{dust}}$) whereas, if dust is ignored, one expects much larger $v_A^{ion}$ speeds in all gas densities above 1 cm$^{-3}$ where the abundance of gaseous ions drastically falls below 1\% \citep{Kim23}. Charged dust therefore allows larger wave intensities for gyro-resonant scattering in the CNM and denser gas phases \citep{Hennebelle23}. 

A quasi-steady state is reached when the growth rate $\Gamma_{\rm grow}$ of the gyro-resonant waves is balanced by the damping rate $\Gamma_{\rm in}$, giving a parallel diffusion coefficient:
\begin{equation}
\kappa_{\parallel}  = \frac{2}{\pi^{3/2}} \frac{c}{e} \Gamma_{\rm in} \ell_{\rm CR} \frac{\sqrt{\rho_i}}{ \sum_j \left( \frac{\alpha_j - 3}{\alpha_j -2} \right) Z_j \, n_j(\mathcal{R}_j > \frac{B}{k})} \; \propto \; T^{1/2} x_i^{1/2} n_{\rm HI}^{3/2} \ell_{\rm CR} .
\label{eq:kappa}
\end{equation}
We note that, under the steady-state self-confinement transport assumption, the diffusion coefficient depends on two opposite trends that limit its potential range inside a cloud since both the gas temperature and ionisation fraction rapidly drop as the gas density increases. We also note that the $\kappa_\parallel/\ell_{\rm CR}$ formulation for the net drift velocity cancels the $\kappa_\parallel$ dependence on the ion Alfvén speed, and therefore also the dependence on the mean magnetic field strength. 

The distance that separates the Reticulum and Eridu filaments ranges from 170~pc to 270~pc given their respective length and distance uncertainty to the Sun. They are also about 150~pc away from other clouds exhibiting \g-ray emissivities close to the average \qlis. Given the 40--60\% change in CR flux between Eridu and the other clouds, we have adopted $\ell_{\rm CR}$=400~pc as a typical CR gradient length scale for the CR energies probed with the LAT. This macroscopic length scale is motivated by differences in CR flux inside various clouds and not by the actual CR gradients that prevail in the immediate vicinity of the Eridu and Reticulum clouds, or those that are induced by magnetic bottlenecks, $v_A^{ion}$ changes, or CR losses in the cloud envelopes \citep{Schlickeiser16,Bustard21}. Given its uncertainty, we explicitly quote the $\ell_{\rm CR}$ dependence in our results.

\begin{figure}[!ht]
  \centering     
  \includegraphics[scale=0.5]{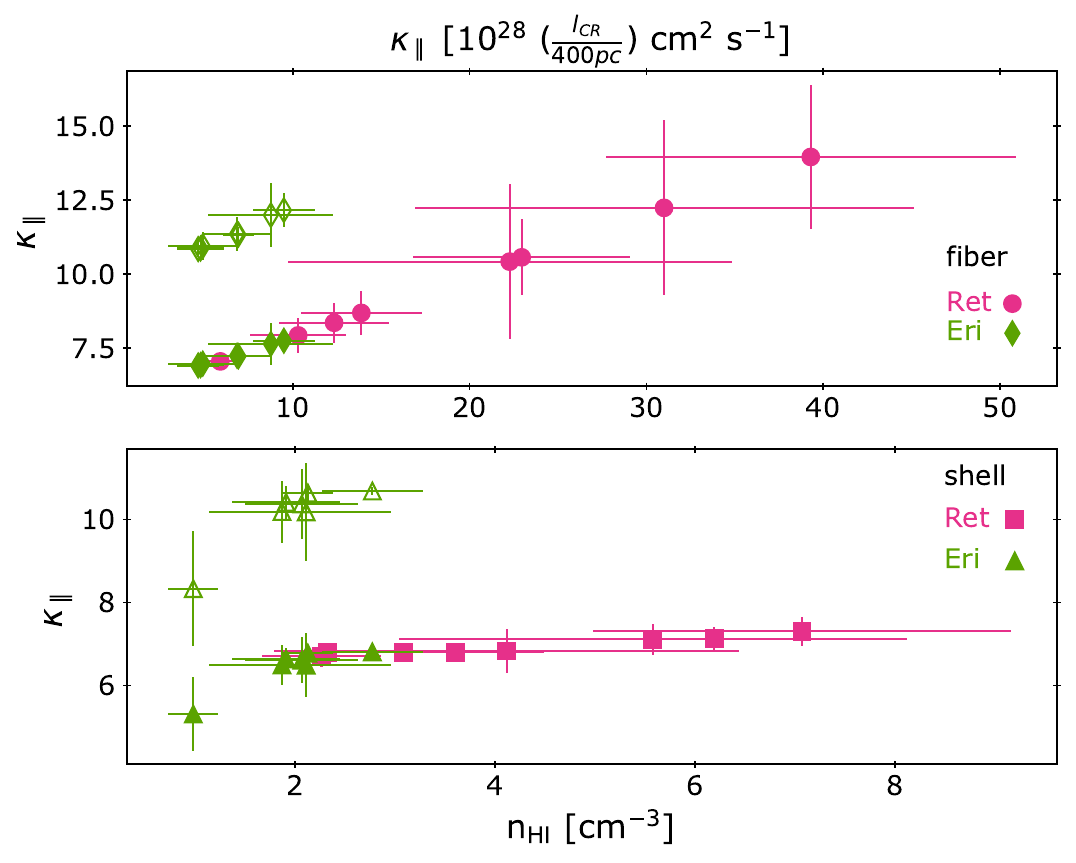}
  \caption{Estimates of the parallel diffusion coefficient in the $1^\circ$-wide squares chosen along the core of the Reticulum (magenta) and Eridu (green) filaments, for the fibre (top) and shell (bottom) geometries, for the same CR flux pervading the two clouds (filled symbols) and for the reduced CR flux in Eridu (open symbols). The data points and error bars respectively give the mean value in each ${\sim}5$~pc-wide square and the rms dispersion at 0.5--0.7~pc scale inside each square. Different $n_{\rm HI}$ and $\kappa_\parallel$ scales are used for each geometry. }  
  \label{fig:KappaCarre}
\end{figure}

Figure \ref{fig:kappaMap} presents maps of the parallel diffusion coefficient obtained for both clouds when using a uniform LOS depth across the whole cloud (the mean one in the fibre configuration or the upper limit in the shell geometry). We further relaxed the uniformity of the LOS depth by calculating the diffusion coefficient in each of the $1^\circ \times 1^\circ$ squares chosen along both filaments, using the fibre LOS depth estimated in each square. The results are shown in Tables \ref{tab:RetISMprop} and \ref{tab:EriISMprop} and as a function of the mean \hi gas density in each square in Fig. \ref{fig:KappaCarre}. For a characteristic CR gradient scale length $\ell_{\rm CR}$ of 400 pc, we find $\kappa_\parallel$ values that typically range from $6\times 10^{28}$ to $14\times 10^{28}$ \cmsqs across the Reticulum cloud in both geometries, and over a smaller range from $5\times 10^{28}$ to $8\times 10^{28}$ \cmsqs for Eridu because of its lower contrast in gas column density. The median $\kappa_\parallel$ value in all the pixels composing the squares along the core of each filament differs by less than 20\% between the two clouds in each geometry. 

The maps show that the diffusion coefficient does not vary substantially as the gas density increases from the cloud envelopes to the cores because the increase in gas density is tempered by the loss in ion abundance and because the damping rise is tempered by the drop in gas temperature and increase in ion mass. The overall variation from edge to core is of the order of two to three. We stress that these $\kappa_\parallel$ estimates are based on the cloud properties sampled at parsec scales rather than at the actual micro-parsec scales of gyro-resonance, so in-situ variations may well be much more severe. 

Figure \ref{fig:kappaMap} illustrates that, when using the same CR flux to excite Alfvén waves in the Reticulum and Eridu clouds, both filaments exhibit very similar diffusion properties over most of their area in each of the fibre or shell configurations. 
The larger $\kappa_\parallel$ values found along the spine of Reticulum in Fig. \ref{fig:KappaCarre} correspond to the densest cores where CRs can flow more rapidly due to the enhanced damping. In the self-confinement theory, CRs create the magnetic fluctuations they interact with, so fewer CRs have less power to generate perturbations to modify their pitch angle. The CR flux reduction in Eridu with respect to Reticulum leads to an increase in diffusion length by the flux ratio of $1.57 \pm 0.09$. Taking this ratio into account, Fig. \ref{fig:kappaMap} shows that CRs should diffuse at about the same speed in the densest parts of Eridu and Reticulum, but twice as fast at the periphery of Eridu than in the envelope of Reticulum, in both geometries. However, the diffusivity difference between the two clouds is too small and the uncertainties in the growth and damping rates are too large to relate the loss of CRs in Eridu to different self-confinement efficiencies.   

Cosmic rays transfer energy to the magnetic fluctuations if they have an anisotropic distribution with a degree of anisotropy larger than $v_A^{\rm ion}/c$, therefore if their drift velocity $v_D$ along the mean magnetic field exceeds $v_A^{\rm ion}$ \citep{Kulsrud69}. If the CRs are well scattered, there is a direct relationship between their anisotropy and spatial density gradient \citep{Zweibel13}, so a difference in CR flux inside Eridu and Reticulum may originate from different CR gradients in their immediate environment (different $\ell_{\rm CR}$ values). Both clouds lie near the dense walls of the Local Valley, but Eridu stands in the vicinity of the Orion-Eridanus superbubble where steep CR gradients are possible even though we have found no evidence for fresh acceleration or reacceleration of CRs in the compressed outer rim of the superbubble \citep{Joubaud20}. Steeper gradients, however, would excite more streaming instabilities and decrease the diffusion coefficient rather than increase it compared to Reticulum. 

On the other hand, Reticulum may be a cooling fragment of an old shell (possibly from a supernova remnant) surrounded by steep CR gradients to efficiently confine the particles inside the shell, but a local boost of the wave growth rate inside this cloud is hard to reconcile with the fact that its CR flux matches that found in other local clouds with different histories and environments. The detection of H$\alpha$ recombination emission towards Reticulum reveals the presence of ionised gas close to the cloud. Efficient streaming and gyro-scattering in the ionised gas may have reduced the CR anisotropy before entering the cloud, thereby preventing the further excitation of scattering waves inside the cloud. But this scenario would lead to faster diffusion in Reticulum than in Eridu since we find no ionised shell around the latter.

We note that $\kappa_\parallel$ linearly scales with $\ell_{\rm CR}$. This is the reason why the present estimates drastically differ from those found by \cite{Armillotta21} for the same gas density. They find very smooth CR densities in the CNM with about two orders of magnitude larger gradient scale lengths in their simulations (Lucia Armillotta, private communication). 

The $\kappa_\parallel$ values we find in these individual clouds in the self-confinement scenario compare well with the range of recent estimates obtained for the entire Milky Way for isotropic diffusion. The latter vary from $4\times 10^{28}$ to $20 \times 10^{28}$~\cmsqs for 10 GeV/n CRs \citep{Johannesson16,Johannesson19,Weinrich20,TorreLuque22}. They are based on direct measurements of CR nuclear spectra measured inside and near the heliosphere (by Voyager, ACE-CRIS, HEAO-3, CREAM, AMS-02, and PAMELA) and on different propagation codes to model spallation reactions (namely GALPROP, USINE, and DRAGON-2). The different estimates vary by a factor five even though the models merely try to constrain an isotropic, homogeneous, environment-independent coefficient in the Galaxy. The convergence between the Galactic averages and the values presented in Fig. \ref{fig:kappaMap} and \ref{fig:KappaCarre} is nice, but it primarily relies on the choice of 400 pc for the $\ell_{\rm CR}$ gradient scale and on the assumption of CR self-confinement in these clouds. As discussed above, variations in the cloud properties (temperature, ionisation fraction, and gas density) tend to compensate each other to reduce the range of $\kappa_\parallel$ values in this transport scenario. 
The choice of $\ell_{\rm CR}$ was not tuned to match Milky Way values, but motivated by the 3D separation between clouds exhibiting different \g-ray emissivities. Nevertheless, while we can compare the diffusion coefficients presented in Fig. \ref{fig:kappaMap} and \ref{fig:KappaCarre} for the two clouds because they rely on the same assumptions and on the same spatial scale and methods to measure the relevant interstellar parameters, we should consider the absolute value of these coefficients with great care because of the unknown state of the plasma and unknown level of CR anisotropy at the gyro-resonance scale. 

\subsection{Velocity dispersion and sonic Mach numbers}
\label{sec:vdisp}

As indicated when opening the discussion section, other CR transport modes relate to the turbulent wandering and bending of magnetic-field lines, in particular in super-Alfvénic turbulence \citep[][and references therein]{Lazarian21,Hu22,Xu22,Sampson23}. 
We therefore explored the turbulence level in gas velocity and in \bsky orientation in each cloud, and we estimated the sonic Mach numbers of the turbulence to gauge the plasma compressibility.

\begin{figure}[!ht]
  \centering     
  \includegraphics[scale=0.45]{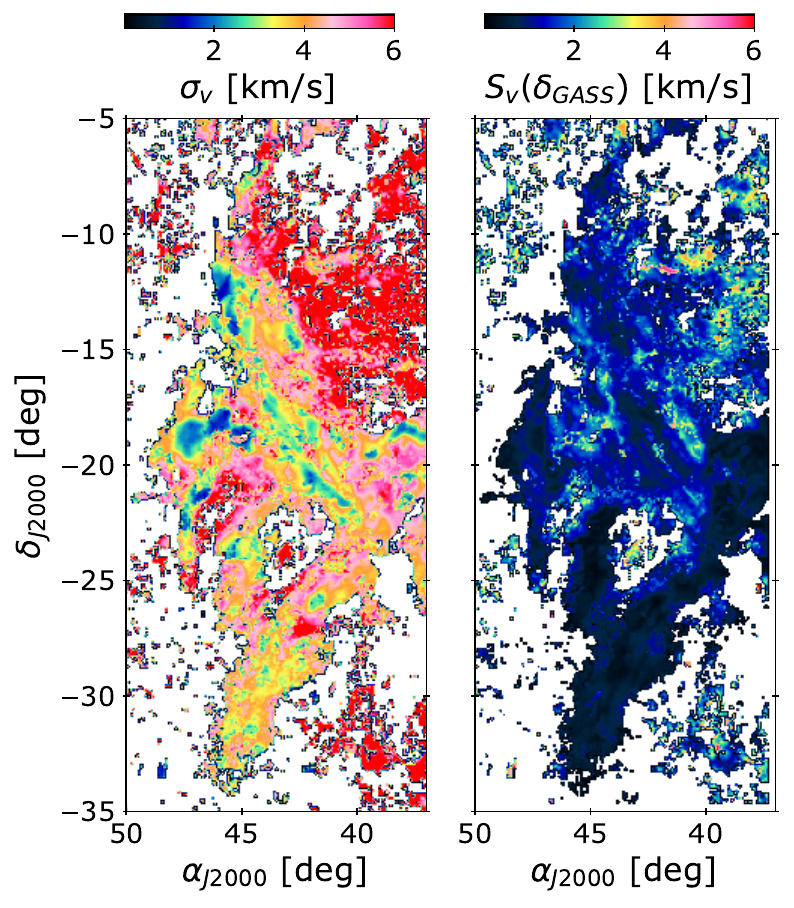}
  \includegraphics[scale=0.45]{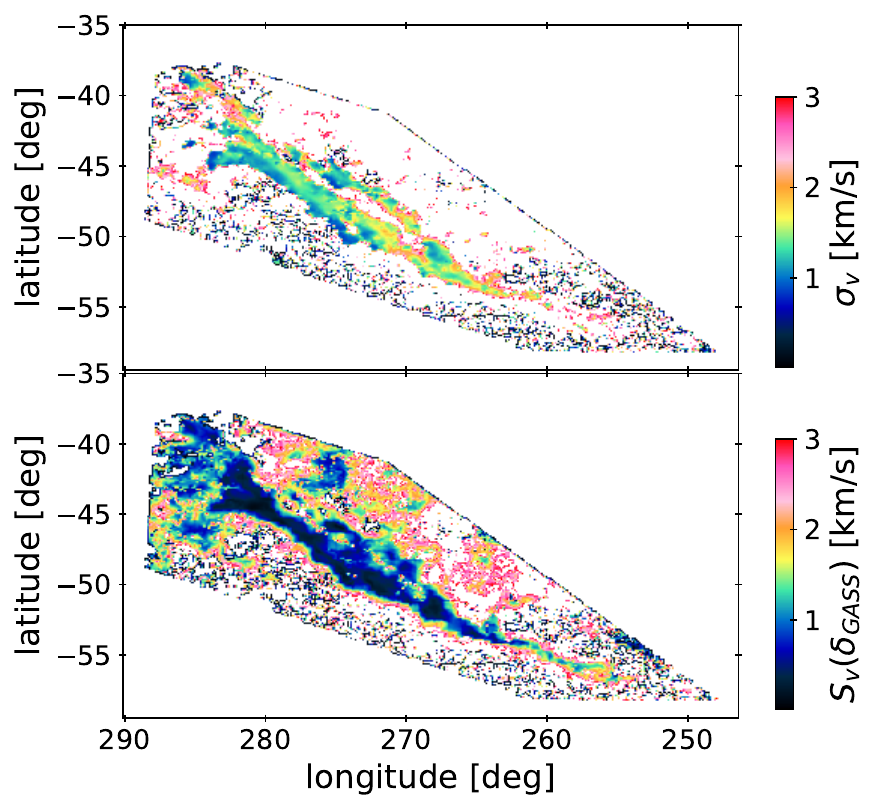}
  \caption{Maps of the velocity line width, $\sigma_v$, and of the velocity dispersion function, $S_v(\delta_{\rm GASS})$, of the main \hi emission line across the Eridu (top) and Reticulum (bottom) clouds. We note the different colour scale for each cloud. }  
  \label{fig:sigV_map}
\end{figure}

We took advantage of the pseudo-Voigt decomposition of the \hi spectra we used for the cloud separation to follow changes in LOS velocity for the main \hi line across each  cloud. Figure \ref{fig:sigV_map} presents the maps of the $\sigma_v$ line widths of the main line across Eridu (top panel) and Reticulum (bottom panel). We also mapped the position-to-position velocity dispersion at the smallest scale resolved by the GASS using a two-point structure function in LOS velocity: 
\begin{equation}
    S_v(\vec{x},\delta)= \sqrt{\frac{1}{N}\sum_{i=1}^N \left[ v_{LOS}(\vec{x}) - v_{LOS}(\vec{x} + \vec{\delta}_i) \right]^2} ,
\end{equation}
where the velocity dispersion compares the line central velocity in a given direction $\vec{x}$ and in a displaced direction $v(\vec{x} + \vec{\delta}_i)$. The average is taken over displaced directions spanning an annulus $\delta/2 \leq \delta_i \leq 3\delta/2$ around the $\vec{x}$ direction, with $\delta$ equal to the effective GASS FWHM angular resolution of 16\farcm2 \citep{Kalberla10}. The resulting $S_v$ maps are shown for the Reticulum and Eridu clouds in the top and bottom panels of Fig. \ref{fig:sigV_map}, respectively.

\begin{figure}[ht]
  \centering     
  \includegraphics[scale=0.5]{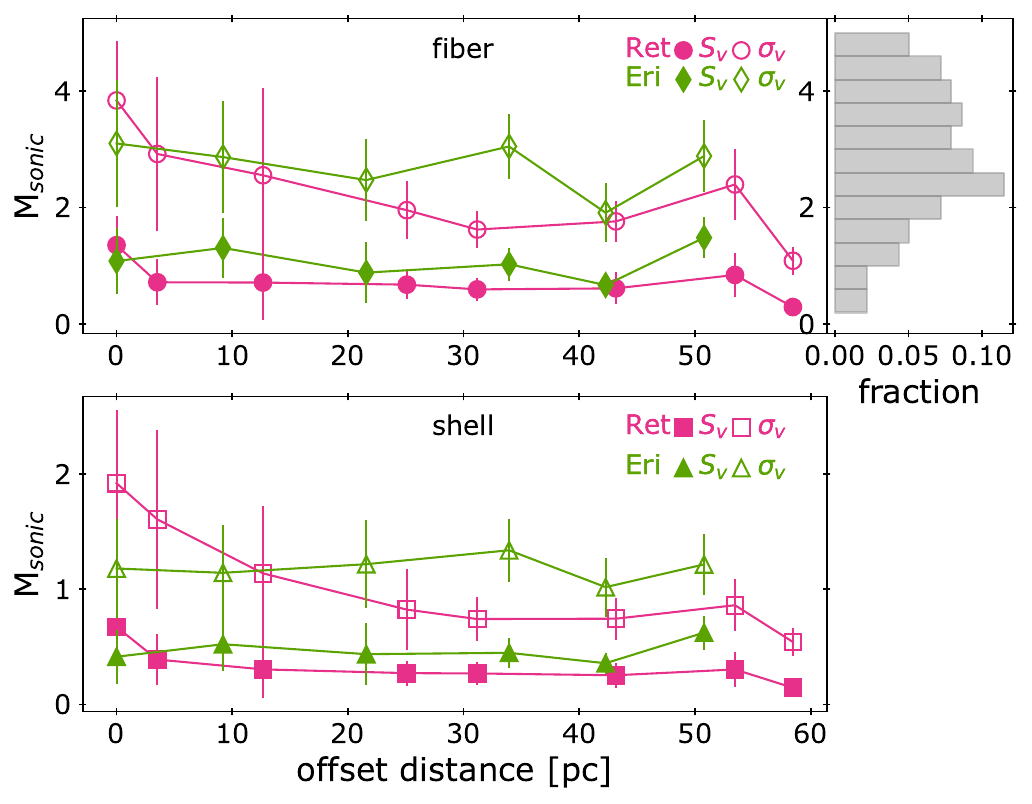}
  \caption{Sonic Mach numbers of turbulent motions in the $1^\circ$-wide squares chosen along the Eridu (green) and Reticulum (magenta) filaments, in the fibre (top) and shell (bottom) geometries, assuming isotropic turbulence and using the $S_v$ dispersion function (dark filled symbols) or the $\sigma_v$ line width (light open symbols) to trace the one-dimensional turbulent velocity dispersion. The data points and error bars respectively give the mean value in each ${\sim}5$~pc-wide square and the rms dispersion at 0.5--0.7~pc scale inside each square. The histogram shows the Mach number distribution in cold CNM clumps from \cite{Heiles03}}  
  \label{fig:Mach_son}
\end{figure}

The $\sigma_v$ line widths are commensurate with the thermal broadening expected from the temperature map inferred in the previous section. The velocity resolution of GASS and the uncertainty in the modelled gas temperature do not allow a reliable subtraction of the thermal broadening to measure the dispersion in turbulent velocities. The $\sigma_v$ line widths therefore provide upper limits to the turbulent motions along the lines of sight. On the other hand, because of the average over the $\delta$ ring, the $S_v$ structure function is closer to a lower limit to the turbulent motions along the line of sight. Both $S_v$ and $\sigma_v$ are one-dimensional measures of the turbulent velocity field. 

Figure \ref{fig:sigV_map} shows that the line width and velocity dispersion values decrease from the cloud edges to the cores of both clouds, as expected from CNM regions. It also indicates that the dispersion in the velocity field of the Eridu cloud is about twice larger than in the more compact Reticulum filament (both in $S_v$ and $\sigma_v$). 
This is expected for a more diffuse cirrus \citep{Audit10}. 

We calculated the sonic Mach numbers of the turbulence in the $1^\circ$-wide squares sampling the core of each filament in the case of isotropic turbulence ($\delta v_{3D}^2 = 3 \delta v_{1D}^2$). Figure \ref{fig:Mach_son} shows that these estimates are consistent with the distribution of Mach numbers obtained in cold CNM clumps by \cite{Heiles03} when comparing the \hi line width with the thermal broadening inferred from spin temperatures. Figure \ref{fig:Mach_son} indicates that the turbulence is mildly supersonic along both Eridu and Reticulum in the fibre configuration and it is trans-sonic in the warmer shell geometry. The Mach numbers compare well in both clouds and the turbulence appears to be moderately compressible in both of them.
% moyenne profil Ret Sv fibre 0.51 shell 0.27
% moyenne profil Ret sigv fibre 1.56 shell 0.83
% moyenne profil Eri Sv fibre 0.79 shell 0.40
% moyenne profil Eri sigv fibre 1.93 shell 1.02

\subsection{Dispersion in the \bsky magnetic field orientation}
\label{sec:Bdisp}

In order to study the geometrical complexity of the magnetic field lines in the Reticulum and Eridu clouds we used the dust polarisation data recorded by \Planck at 353~GHz. We degraded the original 4\farcm8 resolution to an angular resolution of 14\arcmin (FWHM) in order to increase the signal to noise ratio. The data come from the third public data release (PR3), which is described in the Planck 2018 Release Explanatory Supplement\footnote{\url{http://wiki.cosmos.esa.int/planck-legacy-archive}}. The $Q$ and $U$ Stokes parameters provided in the Planck database are defined with respect to the Galactic coordinates, but in a direction opposite to the IAU convention along the longitude axis, so we define the polarisation angle: 
\begin{equation}
    \psi_{IAU} = \frac{1}{2} \arctan(-U,Q) ,
\end{equation}
where $\psi$ is zero at the north Galactic pole and it follows the IAU convention by increasing counter-clockwise with increasing longitude. The $\arctan(\sin, \cos)$ is the arc-tangent function that solves the $\pi$ ambiguity taking into account the sign of the cosine.

The nearby Reticulum and Eridu clouds are the dominant dust clouds along their respective lines of sight, so the dust polarisation measurements give us access to the magnetic field in the plane of the sky, \bsky, in those clouds. In order to quantify the regularity of the \bsky direction, we used the polarisation angle dispersion function \citep{Planck15XIX}:
\begin{equation}
    S_\psi (\vec{x},\delta)= \sqrt{\frac{1}{N}\sum_{i=1}^N \left[ \psi(\vec{x}) - \psi(\vec{x} + \vec{\delta}_i) \right]^2} ,
\end{equation}
where $\Delta \psi_{xi} = \psi(\vec{x})$ - $\psi(\vec{x} + \vec{\delta}_i)$ is the difference between the polarisation angle in a given direction $\vec{x}$ and in a displaced direction $\vec{x}+\vec{\delta}_i$. In practice, we calculated $\Delta \psi_{xi}$ from the Stokes parameters as \citep{Planck15XIX}:
\begin{equation}
    \Delta \psi_{xi} = \frac{1}{2}\arctan(Q_iU_x-Q_xU_i,Q_iQ_x+U_iU_x) ,
\end{equation}
where the $x$ and $i$ indices stand for the central and displaced directions, respectively. The average in $S_\psi$ is taken over displaced directions spanning an annulus around the $\vec{x}$ direction with an inner radius $\delta/2$ and an outer radius $3\delta/2$. We took 15\arcmin~for the $\delta$ lag. 

\begin{figure}[!ht]
  \centering     
  \includegraphics[scale=0.5]{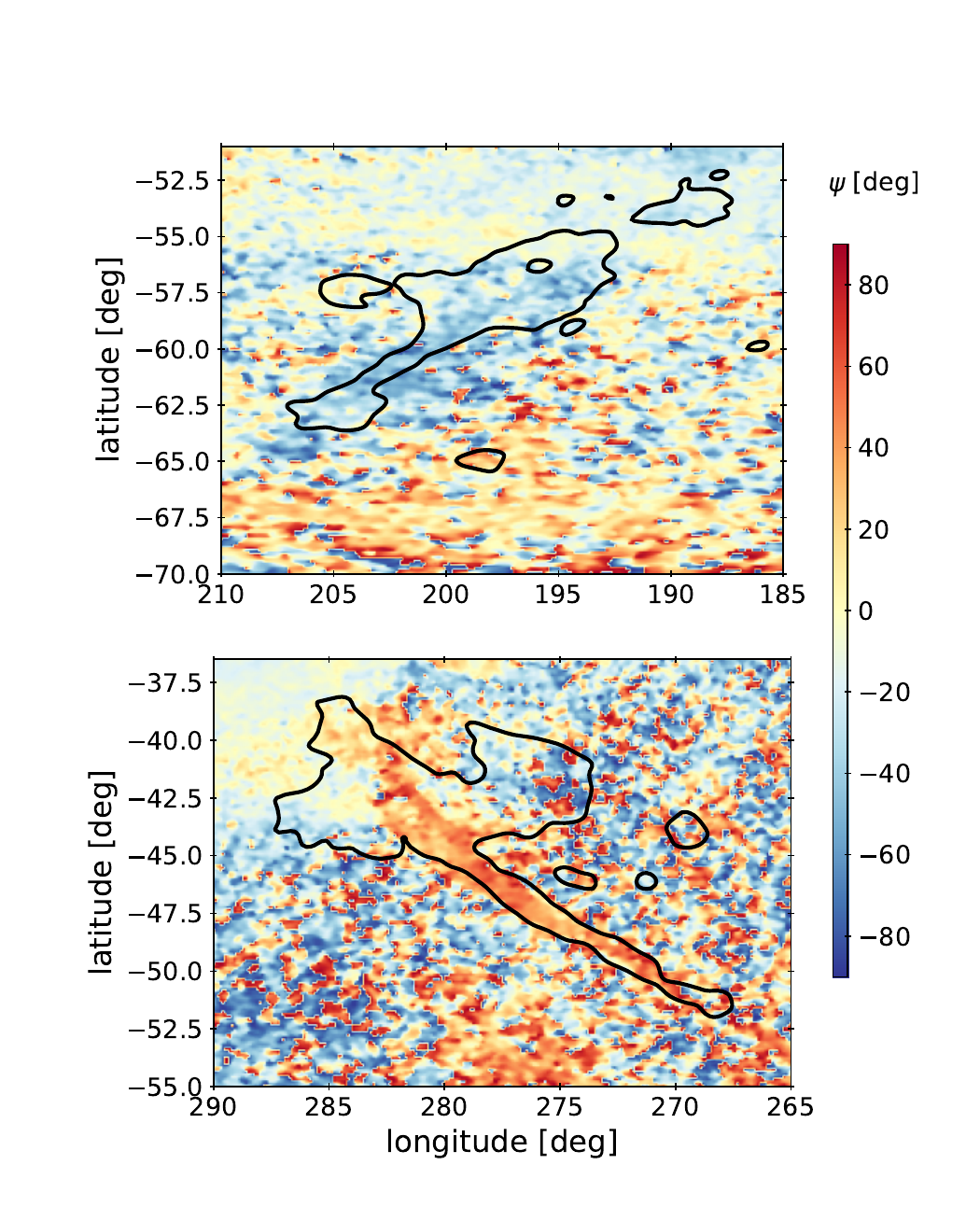}
  \caption{Map of the polarisation angle $\psi$ at 353~GHz with 14\arcmin~resolution in the Eridu (top) and Reticulum (bottom) clouds. The angle gives the orientation of the \bsky field. The black contours delineate the filaments at \nhi column densities of $3 \times 10^{20}$ cm$^{-2}$ and $1.2 \times 10^{20}$ cm$^{-2}$, respectively. }  
  \label{fig:Psi}
\end{figure}

\begin{figure}[!ht]
  \centering     
  \includegraphics[scale=0.48]{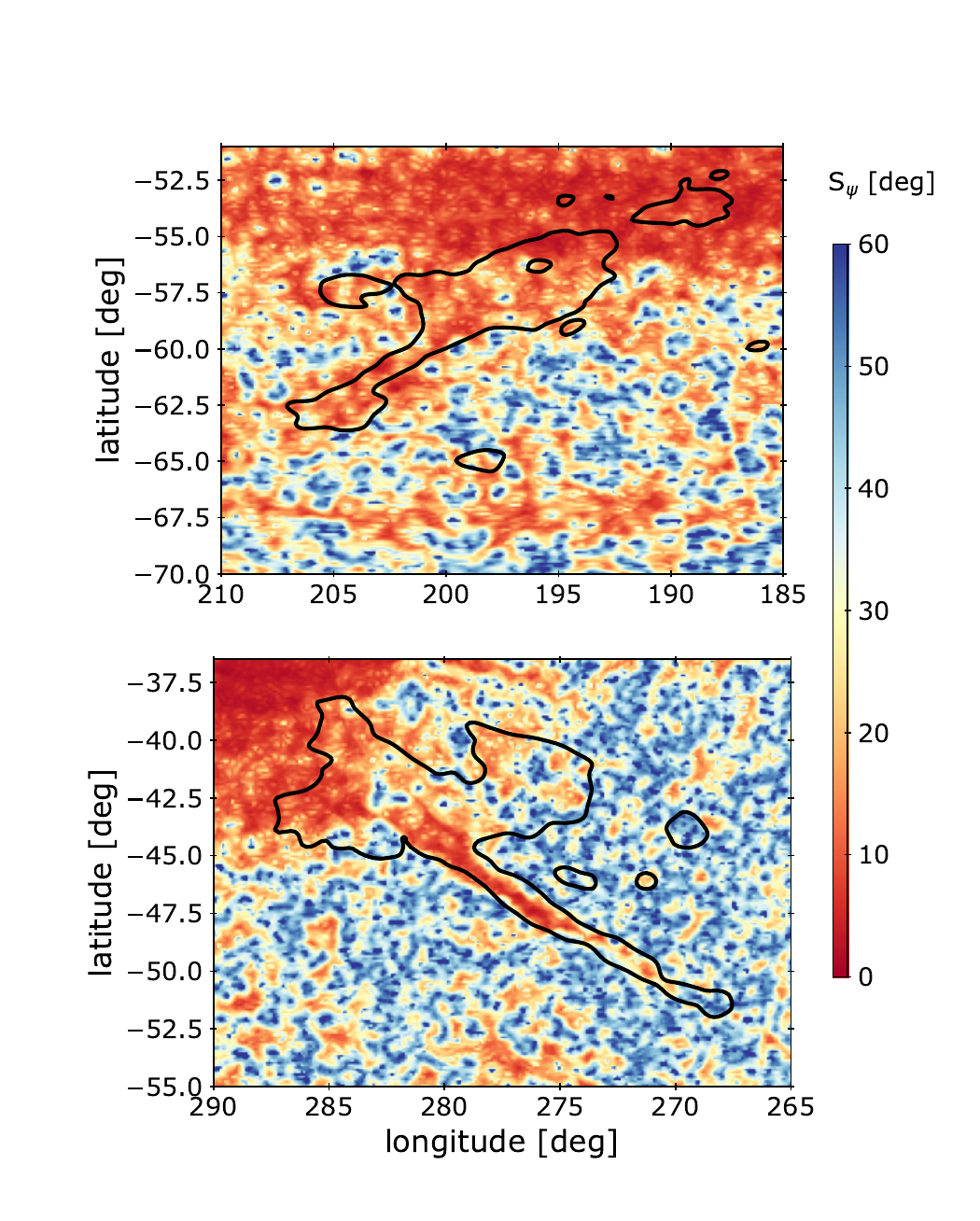}
  \caption{Map in the Eridu (top) and Reticulum (bottom) clouds of the polarisation angle dispersion function $S_\psi$ at 353~GHz with 14\arcmin~resolution and  $\delta$=15\arcmin~lag. The black contours delineate the filaments at \nhi column densities of $3 \times 10^{20}$ cm$^{-2}$ and $1.2 \times 10^{20}$ cm$^{-2}$, respectively.}  
  \label{fig:Sfct}
\end{figure}

The $S_\psi$ dispersion values range from $0^\circ$ for a well ordered \bsky magnetic field to $S_\psi= 90^\circ$ for a disordered one. When the polarisation signal is dominated by noise, $S_\psi$ converges to $\pi/ \sqrt{12}$ ($\approx$ 52$^\circ$). The results on the \bsky field orientation and on its dispersion are displayed as field lines in Fig. \ref{fig:Bskymaps}, as $\psi$ angle maps in Fig. \ref{fig:Psi}, and as $S_\psi$ angle dispersion maps in Fig. \ref{fig:Sfct}. 

Atomic filamentary clouds have mean magnetic fields often aligned with the filament axis \citep{Clark14,Clark19} whereas the field orientation turns to perpendicular or to disorder in the dense molecular parts \citep{Soler13,Planck16XXXV}. The atomic Eridu and Reticulum clouds follow the expected trend.
Figures \ref{fig:Bskymaps}, \ref{fig:Psi}, and \ref{fig:Sfct} jointly show that the \bsky magnetic field is well ordered along the axis of both filaments, with a stable direction along nearly their whole length, and with a dispersion lower than 10$^\circ$ on a 1-pc scale over large parts of the filament extent. Moreover, these ordered fields are almost perpendicular to the orientation of the Galactic magnetic field inferred in the solar neighbourhood from dust polarisation in local clouds \citep{Planck16XLIV} or by the field draping around the heliosphere \citep{Zirnstein16}.

The Reticulum field appears to be slightly more ordered in its dense part than in Eridu, likely because of the compression that originally formed the compact fibre or shell. The dispersion in field orientation appears to be rather uniform across the whole of Eridu and is a little more contrasted between the filament axis and edges of Reticulum. The parsec-scale data, however, do not suggest that CRs have slowed down in Reticulum because of a much more tangled field. To the contrary, compared to Eridu, they should flow along more ordered fields over a large fraction of the filament length. 

Because of the lateral super-diffusion of turbulent magnetic field lines, CRs with large enough pitch angles can bounce back on magnetic mirrors and leap onto adjacent lines \citep[see Fig. 1 of][]{Lazarian21} whereas the low pitch-angle CRs (in the loss cone for mirroring) keep flowing on and undergo gyro-resonant pitch-angle scattering. Mirror diffusion efficiently slows down the effective CR speed along the mean field, but only for the small fraction of particles with large pitch angles, so the overall reduction in $\kappa_\parallel$ is small. It is typically less than a factor of 2 and fades out for CR energies below 100 GeV \citep{Lazarian21}. The fact that we find comparable dispersion levels in \bsky orientation in the two clouds does not suggest widely different levels of mirror diffusion, unless the field gets much more tangled in the more numerous CNM clumps of Reticulum at unresolved scales below one parsec. We are investigating this possibility with interstellar MHD simulations.

\subsection{Magnetic-field strength}
\label{sec:Bsky}

\begin{figure}[!ht]
  \centering     
  \includegraphics[scale=0.5]{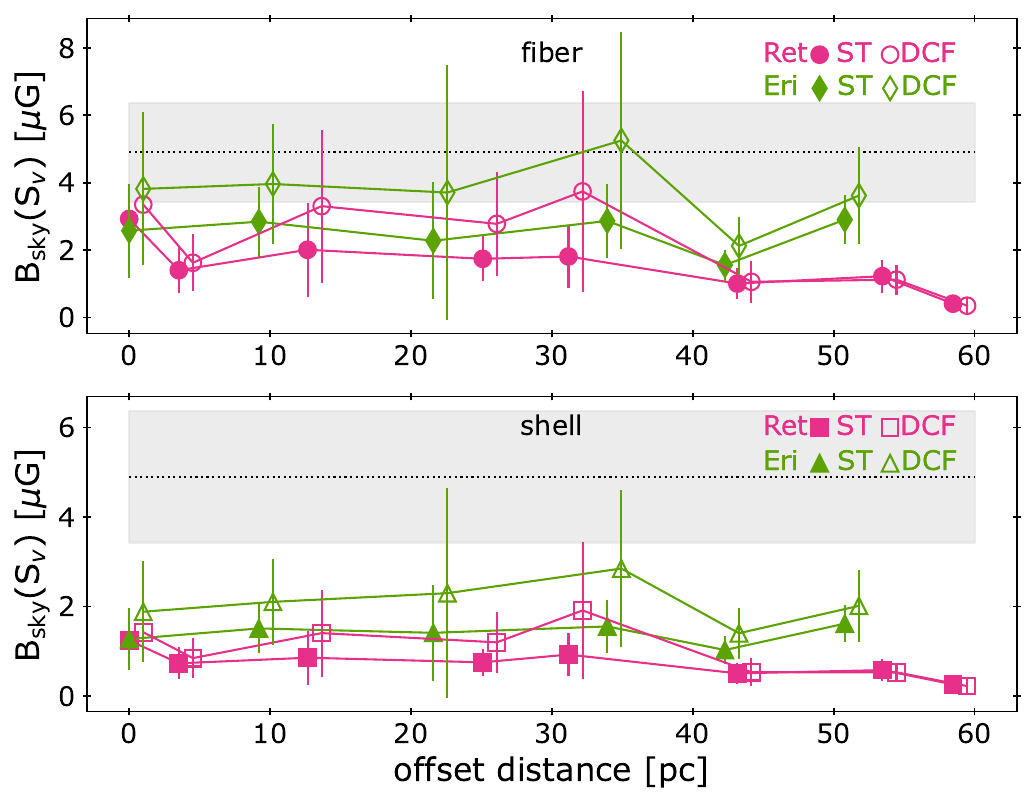}
  \caption{Plane-of-sky magnetic-field strengths in the $1^\circ$-wide squares chosen along the Reticulum (magenta) and Eridu (green) filaments, in the fibre (top) and shell (bottom) geometries, using the $S_v$ dispersion function in the ST (dark filled symbols) or DCF method (light open symbols). The data points and error bars respectively give the mean value in each ${\sim}5$~pc-wide square and the rms dispersion at 0.5--0.7~pc scale inside each square. The offset distance of the DCF data points has been slightly enlarged for clarity. The grey band shows the mean CNM \bsky value from \cite{Heiles05a} assuming $B_{\rm sky}^2 = \frac{2}{3} B^2$. }
  \label{fig:Bsky}
\end{figure}

The total magnetic field is composed of a mean component, $\vec{B_0}$, and a small fluctuating one, $|\vec{\delta B}|\ll |\vec{B_0}|$, so that the magnetic energy density is $\frac{B^2}{8\pi} = \frac{1}{8\pi}[B_0^2+\delta B^2+2\vec{\delta B}.\vec{B_0}]$. To infer the mean field strength from dust polarisation, the Davis-Chandrasekhar-Fermi (DCF) method assumes that isotropic turbulent motions initiate incompressible, transverse, Alfvén waves ($\delta \vec{B} \cdot \vec{B}_0 = 0$) in the frozen field \citep{Davis51,Chandrasekhar53}. 
They also assume that the kinetic energy of the turbulence is equal to the fluctuating magnetic energy density, and that the dispersion in polarisation angle $\psi$ traces the dispersion in magnetic-field orientation. From the equipartition condition, $\frac{1}{2}\rho \delta v^2 = \frac{\delta B^2}{8\pi}$ and from $S_\psi \approx \delta B / B_0$ one obtains:
\begin{equation}
    B_{\rm sky}^{\rm DCF} = f \sqrt{4\pi \rho} \frac{\delta v}{S_\psi} ,
    \label{eq:B_DCF}
\end{equation}
where the $f$ correction factor varies with authors between 0.3 and 0.7, for instance $f = 1/\sqrt{3}$ for isotropic turbulence \citep{Chandrasekhar53}. Other values are reviewed by \citet[][Sect. 2.2.3]{Skalidis21}. We adopted their $f=0.5$ recommendation for this work. 

The DCF method tends to overestimate the actual field strength because of line-of-sight and beam averaging in the data and because the method overlooks the strong anisotropy in ISM turbulence and the role of compressible modes in the cross-term $\vec{\delta B}.\vec{B_0}\neq0$ \citep{Skalidis21}. To include the compressible modes, these authors propose an alternative formulation that agrees within 20-30\% with MHD simulation data: 
\begin{equation}
    B_{\rm sky}^{\rm ST} =  \sqrt{2\pi \rho} \frac{\delta v}{\sqrt{S_\psi}}
    \label{eq:B_ST} .
\end{equation}

We applied both methods to calculate the plane-of-sky magnetic-field strengths in the 1$^\circ$-wide squares chosen along the Reticulum and Eridu filaments, with the velocity dispersion function $S_v$ as a proxy for the turbulent velocity $\delta v$. The results are shown in Fig. \ref{fig:Bsky} and in Tables \ref{tab:RetISMprop} and \ref{tab:EriISMprop}. Following the discussion by \cite{Skalidis21}, the $B_{\rm sky}^{\rm ST}$ and $B_{\rm sky}^{\rm DCF}$ estimates likely bracket the true \bsky value, the latter being about 40\% (50\%) larger than the former in the case of Reticulum (Eridu). 
We also calculated upper limits to \bsky using the $\sigma_v$ line width as an upper limit to the turbulent velocity. The results are provided in Tables \ref{tab:RetISMprop} and \ref{tab:EriISMprop}. They are about three times larger than the $S_v$-based estimates for each method.

\begin{figure*}[!ht]
  \centering     
  \includegraphics[width=0.75\hsize]{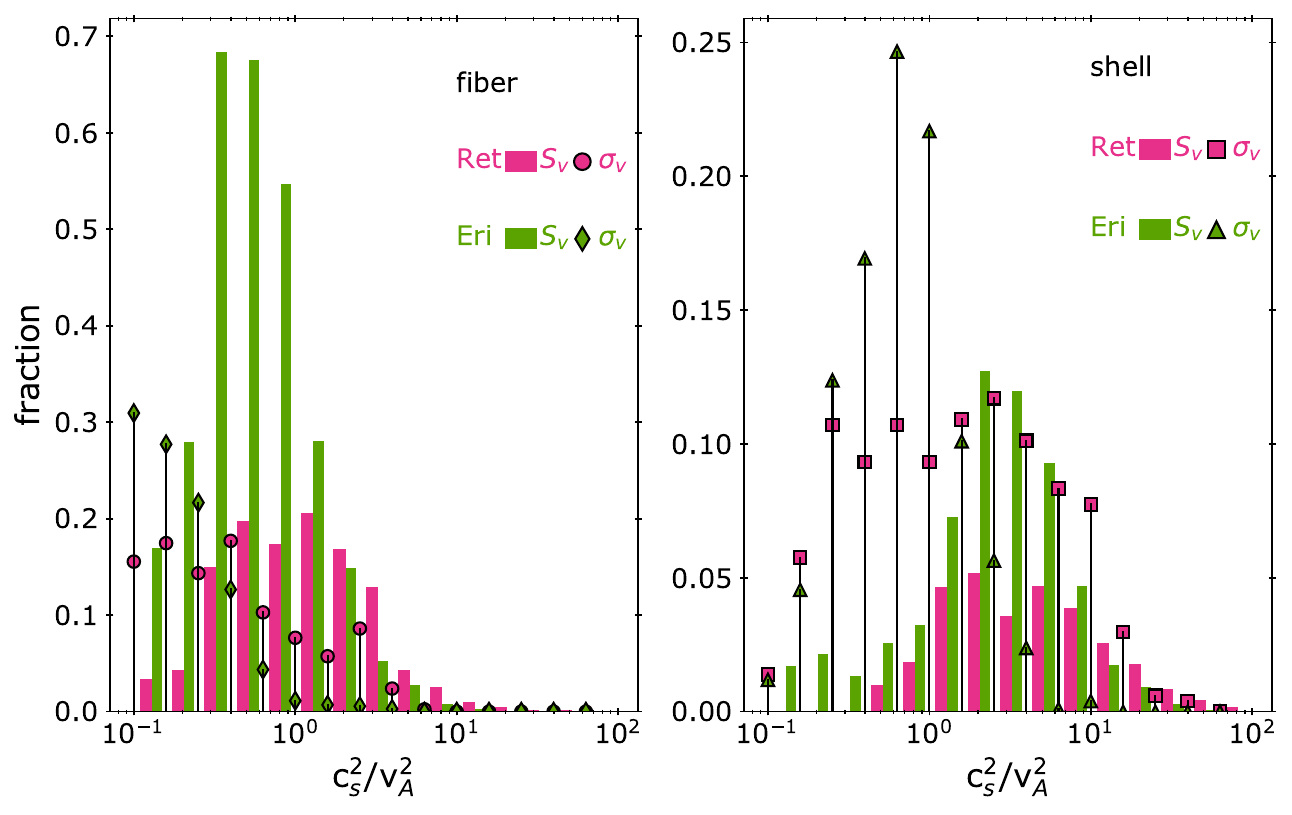}
%  \sidecaption
  \caption{Histograms of the gas magnetisation levels in the $1^\circ$-wide squares chosen along the Eridu (green) and Reticulum (magenta) filaments, for each geometry, and using the $S_v$ velocity dispersion (filled bars) or the $\sigma_v$ line width (markers) in the ST formulation to infer \bsky and the total magnetic-field strength. } 
  \label{fig:betamgn}
\end{figure*}

Figure \ref{fig:Bsky} shows that the \bsky field strength is rather uniform along the core of each cloud and that it is about one and a half time larger along Eridu than along Reticulum: around 3.1 (1.7) $\mu$G for Eridu and 1.9 (0.9) $\mu$G for Reticulum in the fibre (shell) configuration if we use the $S_v$ velocity dispersion to trace the turbulent motions. Three times larger upper limits to the fields are found when using the $\sigma_v$ line width instead. The difference between Eridu and Reticulum does not depend much on the cloud geometry, but one should keep in mind that the magnetic-field component along the line of sight may significantly differ in both clouds. Figure \ref{fig:Bsky} also illustrates that the \bsky strengths in Eridu and Reticulum compare reasonably with the median total field $B_{tot} = 6 \pm 1.8 \; \mu$G inferred from Zeeman splitting of \hi lines and Monte Carlo simulations \citep{Heiles05a} if one assumes $B_{\rm sky}^2 = (2/3) \, B^2$ for a random field orientation with respect to the line of sight \citep{Heiles05a}. 

We cannot estimate whether the MHD turbulence is super-Alfvénic or not since both \bsky derivations assume equipartition between the fluctuating kinetic and magnetic energy densities, whether the latter is dominated by $\delta B^2/8\pi$ in the case of incompressible turbulence (DCF) or by $\delta B B / 4\pi$ when compressible modes are present (ST). In both cases, the Alfvén Mach numbers will be close to one. The large-scale order we observe in the \bsky orientation in both clouds suggests sub- or trans-Alfvénic turbulence as often found for the atomic gas phases in interstellar MHD simulations. 

We can estimate the gas magnetisation level from the squared ratio of the sound to Alfvén speed, $c_s^2/v_A^2$ \citep{Soler13}, which is close to the gas to magnetic pressure ratio. To derive the Alfvén velocity of the total gas, we assumed $B^2 = (3/2) B_{\rm sky}^2$ and the ST formulation for \bsky. Figure \ref{fig:betamgn} shows the normalised distributions of the ratios found in the $1^\circ$-wide squares sampling the gas along the core of both filaments. If we use the $B_{\rm sky}(S_v)$ estimate, the gas and magnetic pressures compare well across both clouds in their fibre geometry and the gas pressure slightly exceeds the magnetic one in the shell configuration. The magnetisation is skewed to values below one (i.e. to dominant magnetic pressure) when using the $B_{\rm sky}(\sigma_v)$ upper limits to the field, but not severely, so equivalent gas and magnetic pressures may prevail over large parts of the clouds. Therefore, whether in terms of \bsky orientation dispersion or in terms of magnetic stiffness against turbulent gas motions, we find no important difference between the two clouds at a parsec scale that would suggest more magnetic-field tangling and slower CR diffusion in one of them. 

\section{Conclusions}
We compared the CR flux in the atomic phase of two nearby clouds that have many properties in common. They are both elongated filaments, which cross the same hollow sector of the Local interstellar Valley at comparable distances from the Sun. Both are threaded by ordered magnetic fields that are mostly aligned with the long axis of the filament and that are largely inclined with respect to the Galactic plane and to the large-scale magnetic-field direction in the local ISM. Both clouds have magnetic pressures of the same magnitude as the thermal ones. Nevertheless, we find that the \g-ray emissivity per gas nucleon in one cloud, Reticulum, is $1.57 \pm 0.09$ times larger than in the other cloud, Eridu, over the whole energy range of analysis, from 0.16 to 63 GeV. The difference cannot be attributed to uncertainties in gas mass because the correlation between the dust and \g-ray data allows an efficient detection of any additional gas that is not traced by \hi line emission (as we find in Reticulum) and the \g-ray contribution of this additional gas is included in the model. Furthermore, we cannot reduce the CR flux ratio below $1.32 \pm 0.08$ when allowing for strong differences in \hi spin temperature between the two clouds (at the expense of a poorer \g-ray fit). 

Beyond characterising the global magnetic orientation and average field strength of a few microGauss in the clouds, we studied turbulence characteristics at a parsec scale that might hint at different environmental conditions for CR diffusion, namely the dispersion in gas velocity and in magnetic-field orientation, differences in the sonic Mach numbers (compressibility) of the turbulence, and differences in magnetisation level (magnetic stiffness against gas turbulent motions). We find that the two clouds compare well in all these aspects. The only notable difference is that the core of the Reticulum cloud has collapsed (or has been compressed) to larger densities typical of the CNM and DNM phases, whereas a large fraction of the more diffuse gas in Eridu may be transitioning through the unstable, lukewarm phase. MHD simulations will help in studying the level of magnetic-field tangling in the different atomic and DNM gas phases, or in shell compression behind a slow shock, in order to search for changes that may impact the collective drift of the CR population, such as changes in the turbulent energy fraction of fast compressive modes or in the density of magnetic mirrors \citep{Lazarian23}. 

Alternatively, the $1-500$ GeV/n CRs we probed in \g rays may primarily scatter off MHD waves they have excited through streaming instability. If so, one expects a stronger damping in the denser Reticulum cloud, but we show that the difference in expected steady-state diffusion coefficient between the two clouds can be small (less than a factor of two) because the increase in diffusion length with the ambient gas density is moderated by the corresponding drop in gas temperature and ionisation fraction. We plan to further investigate the impact of charged dust grains, which should reduce the damping rates along the core of the two clouds \citep{Hennebelle23}. However, we only probed the cloud properties at a parsec scale, whereas the growth and damping of the relevant MHD waves occur at wavelengths of a few micro-parsecs where the interstellar gas properties cannot be explored. Moreover, in the self-confinement mode, the wave growth rate in each cloud relates to the degree of  anisotropy in CR pitch angle or, equivalently, to the steepness of the local CR spatial density gradient \citep{Zweibel13}, which could differ in the vicinity of the clouds or in bottlenecks along their edge (different $\ell_{\rm CR}$). 

The damping rates we estimate suggest that the magnetic perturbations propagating along the field lines should dissipate over typical lengths of tens to hundreds of micro-parsecs \citep[$L \approx v_A^{ion}/\Gamma_{in}$,][]{Spangler11}. The median of their distribution is twice as large in Eridu as in Reticulum. The dissipation lengths correspond to proton gyro-radii for energies in the ${\sim}$10 GeV to ${\sim}$1 TeV band, with a median energy three times larger in Eridu than in Reticulum. The waves must be constantly replenished or we would observe a spectral change in \g-ray emissivity in the corresponding ${\sim}$1 to ${\sim}$100 GeV band as the higher-energy particles would leak out more rapidly. The fact that we find a change in CR flux but not in spectrum between the two clouds is therefore an important constraint. The same overall spectrum has been observed in several clouds in the local ISM \citep{Grenier15}, but with improved photon statistics the \Fermi-LAT data can now serve to precisely test the uniformity of this spectrum in different types of atomic clouds and independently of the CR flux pervading them.

%------------acknowledgements ---------------------------------
%---------------------------------------------------------------------------
\begin{acknowledgements}
We thank %the referee for an interesting discussion and 
Jeong-Gyu Kim for sharing his interstellar model for the solar neighbourhood conditions, Lucia Armillotta for sharing information from her work, the members of the Interstellar Institute (I2) CNRS International Research Network for enlightening discussions, and the referee for helping us clarify several points in the discussion. We acknowledge the financial support by the LabEx UnivEarthS (ANR-10-LABX-0023 and ANR-18-IDEX-0001) for this work.
The \textit{Fermi} LAT Collaboration acknowledges generous ongoing support
from a number of agencies and institutes that have supported both the
development and the operation of the LAT as well as scientific data analysis.
These include the National Aeronautics and Space Administration and the
Department of Energy in the United States, the Commissariat \`a l'Energie Atomique
and the Centre National de la Recherche Scientifique / Institut National de Physique
Nucl\'eaire et de Physique des Particules in France, the Agenzia Spaziale Italiana
and the Istituto Nazionale di Fisica Nucleare in Italy, the Ministry of Education,
Culture, Sports, Science and Technology (MEXT), High Energy Accelerator Research
Organization (KEK) and Japan Aerospace Exploration Agency (JAXA) in Japan, and
the K.~A.~Wallenberg Foundation, the Swedish Research Council and the
Swedish National Space Board in Sweden. Additional support for science analysis during the operations phase is gratefully acknowledged from the Istituto Nazionale di Astrofisica in Italy and the Centre National d'\'Etudes Spatiales in France.
\end{acknowledgements}

%------------bibliography ---------------------------------
\bibliographystyle{aa}
\bibliography{biblio}

%------------Annex ---------------------------------
\appendix

\section{Distance estimates to the nearby clouds} 
\label{sec:dist}
\begin{figure*}[!ht]
  \centering     
  \includegraphics[scale=0.7]{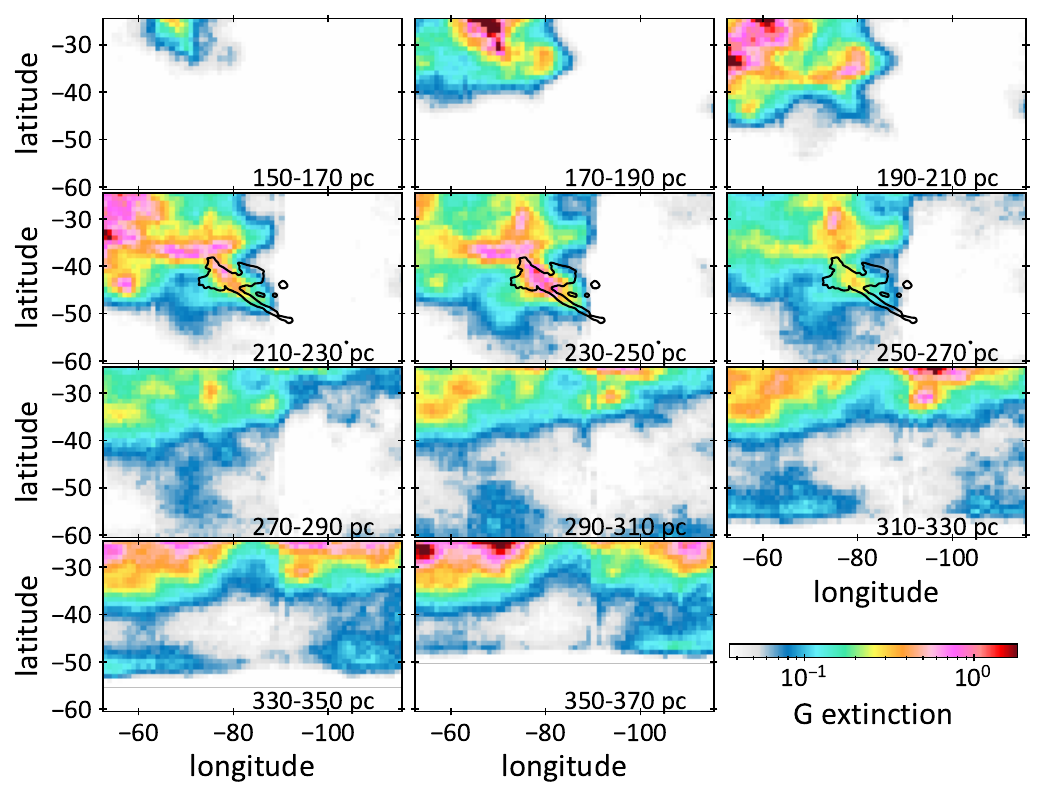}
  \caption{ Maps of the G-band extinction optical depth integrated over 20 pc-wide intervals in  distance from the Sun across the analysis region. The data come from the 3D distribution of \cite{Leike20}. The distance ranges are labelled in each map. The black contours delineate the Reticulum \hi cloud as in Fig. \ref{fig:Bskymaps}. }  
  \label{fig:dist_Ret}
\end{figure*}
\begin{figure*}[!ht]
  \centering     
  \includegraphics[scale=0.7]{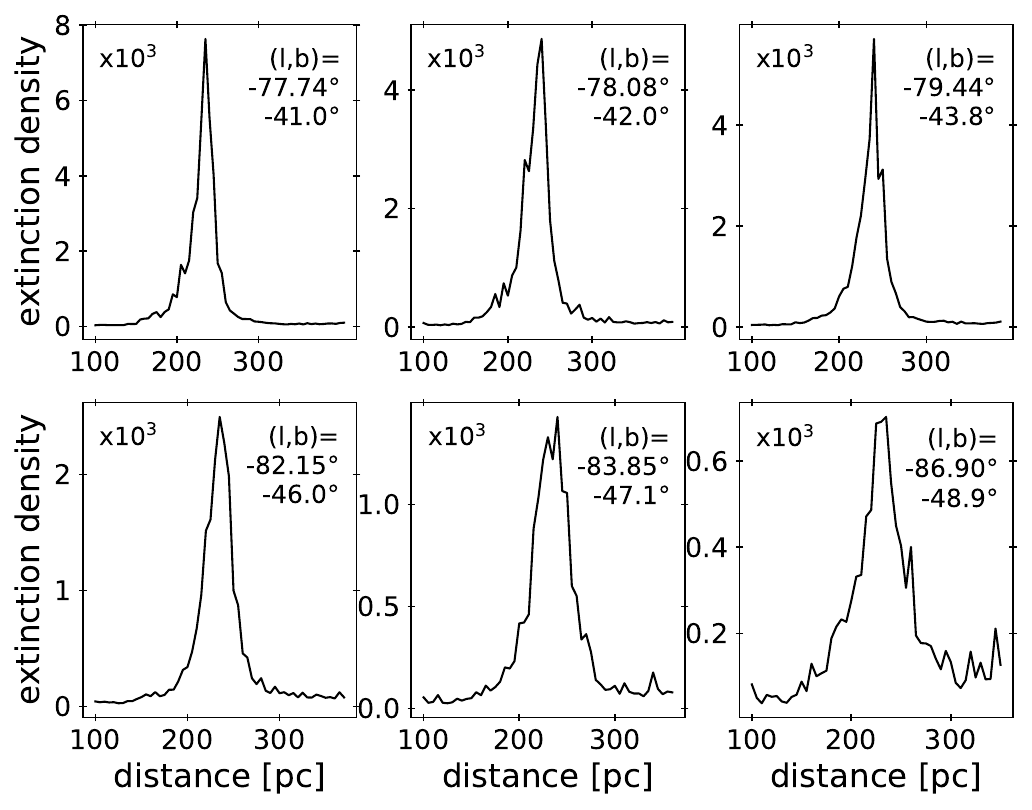}
  \caption{ Dust extinction density profiles as a function of radial distance from the Sun for selected lines of sight along the Reticulum filament. The data come from the 3D distribution of \cite{Leike20}. }  
  \label{fig:distprof_Ret}
\end{figure*}
\begin{figure*}[!ht]
  \centering     
  \includegraphics[scale=0.7]{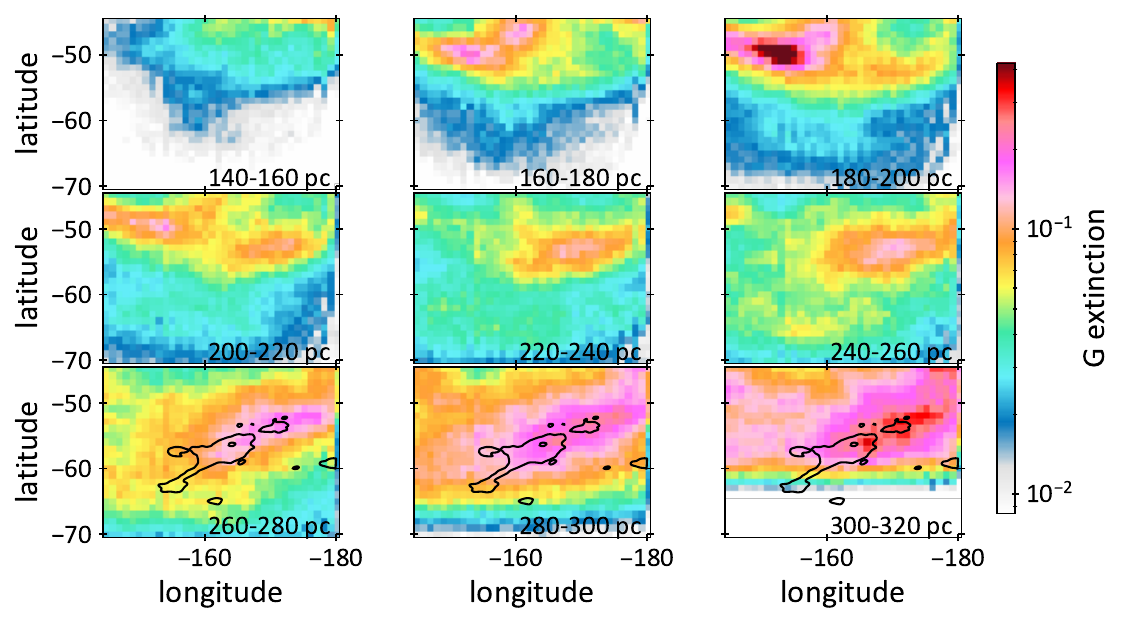}
  \caption{ Maps of the G-band extinction optical depth integrated over 20 pc-wide intervals in distance from the Sun, towards the Orion-Eridanus region. The data come from the 3D distribution of \cite{Leike20}. The dust cloud that spans the region nearly horizontally around $b=50^\circ$ and at distances between 160 and 260 pc marks the rim of the Orion-Eridanus superbubble. At larger distances, the black contours delineate the Eridu \hi cloud as in Fig. \ref{fig:Bskymaps}. The distance ranges are labelled in each map. }  
  \label{fig:dist_Eri}
\end{figure*}

We used the 3D dust distribution reconstructed from stellar surveys by \cite{Leike20} out to distances of about 400 pc from the Sun to place constraints on the distance to different cloud complexes seen in the analysis region. Figure \ref{fig:dist_Ret} displays the dust extinction in the G band in contiguous 20-pc-wide intervals. The lack of dust information at latitudes below about 50$^\circ$ for the three most distant intervals is due to the limited size of the 3D data cube perpendicular to the Galactic plane ($\pm 270$~pc). In addition to those maps, one can plot the extinction profile as a function of distance along sight lines of large \nh column density in each cloud. Examples are shown in Fig. \ref{fig:distprof_Ret} when probing directions along the Reticulum filament. We used the peaks found in the dust profiles and the correspondence in shape found between atomic clouds in Fig. \ref{fig:NHI} and dust clouds in Fig. \ref{fig:dist_Ret} to estimate the cloud distances that are given in Sect. \ref{sec:dist}. 

We repeated the analysis to revisit the distance to the Eridu cloud. The latter, being rather diffuse, is more difficult to identify in the dust data, but Fig. \ref{fig:dist_Eri} suggests that it lies at about 300 pc from the Sun, beyond the edge of the Orion-Eridanus superbubble that is clearly detected at closer distances around 200 pc as the long arc crossing the maps horizontally at latitudes around -50$^\circ$.

\section{ISM properties in the 1$^\circ$-wide squares along the cloud filaments}

Tables \ref{tab:RetISMprop} and \ref{tab:EriISMprop} list the mean value and rms dispersion of the interstellar properties found in the $1^\circ \times 1^\circ$ squares that are depicted in Fig. \ref{fig:CloudWidth} and that span the length of the Reticulum and Eridu filaments, along their core. The properties include the LOS depth, the \nhi column density and $n_{\rm HI}$ volume density in the atomic gas, the \hi velocity dispersion in the plane of the sky ($S_v$) and along the line of sight ($\sigma_v$ line width), the gas temperature $T$ and the abundance ($x_{\rm ion}$) and atomic weight ($A_{\rm ion}$) of ions inferred from the gas density and the \cite{Kim23} model, the \bsky magnetic-field strengths estimated according to Eqs. \ref{eq:B_DCF} and \ref{eq:B_ST} with the $S_v$ velocity dispersion or $\sigma_v$ line width as a proxy for the turbulent velocity $\delta v$, the sonic Mach numbers $M_S$ of the turbulence for the same options in $\delta v$, the Alfvénic Mach numbers $M_A$ of the MHD turbulence inferred with the ST and DCF methods, the ion--neutral damping rate according to Eq. \ref{eq:Soler}, and the steady-state $\kappa_\parallel$ diffusion coefficient for self-confinement with the CR flux reaching the heliosphere in the 1--20 GV rigidity range.  

% ********************************************************
\begin{table*}
\caption{ISM properties along the Reticulum filament, for the fiber (regular font) or shell (bold font) geometry. The values and errors respectively give the mean value in each ${\sim}5$~pc-wide square and the rms dispersion at 0.5--0.7~pc scale within each square. The Galactic coordinates refer to the square centre.}
\centering
\scalebox{0.85}{
\renewcommand{\arraystretch}{2}
\begin{tabular}{ccccccccc}
\hline
  
long. [deg] & 282.09 & 281.58 & 280.05 & 277.51 & 276.15 & 272.59 & 269.88 & 268.18 \\ 
lat. [deg] & -41.25 & -42.0 & -43.88 & -46.25 & -47.38 & -49.12 & -50.88 & -51.5 \\ \hline 
$N_{\rm HI}$ [$10^{20}$~cm$^{-2}$] & 4.4 & 3.8 & 3.4 & 2.5 & 2.2 & 1.9 & 1.4 & 1.4 \\ \hline 
LOS depth [pc] & 3.6 $\pm$ 0.5 & 5.4 $\pm$ 0.7 & 3.6 $\pm$ 0.4 & 3.7 $\pm$ 0.5 & 5.2 $\pm$ 0.6 & 5.0 $\pm$ 0.6 & 4.4 $\pm$ 0.6 & 7.9 $\pm$ 1.0 \\ 
 & \textbf{20.0} $\pm$ \textbf{2.5} & \textbf{20.0} $\pm$ \textbf{2.5} & \textbf{20.0} $\pm$ \textbf{2.5} & \textbf{20.0} $\pm$ \textbf{2.5} & \textbf{20.0} $\pm$ \textbf{2.5} & \textbf{20.0} $\pm$ \textbf{2.5} & \textbf{20.0} $\pm$ \textbf{2.5} & \textbf{20.0} $\pm$ \textbf{2.5} \\ \hline 
$n_{\rm HI}$ [cm$^{-3}$] & 39 $\pm$ 11 & 23 $\pm$ 6 & 31 $\pm$ 14 & 22 $\pm$ 13 & 14 $\pm$ 3 & 12 $\pm$ 3 & 10 $\pm$ 3 & 6 $\pm$ 1 \\
 & \textbf{7} $\pm$ \textbf{2} & \textbf{6} $\pm$ \textbf{2} & \textbf{6} $\pm$ \textbf{3} & \textbf{4} $\pm$ \textbf{2} & \textbf{4} $\pm$ \textbf{1} & \textbf{3} $\pm$ \textbf{0.8} & \textbf{2} $\pm$ \textbf{0.5} & \textbf{2} $\pm$ \textbf{0.4} \\ \hline 
$S_v$ [km~s$^{-1}$] & 0.78 & 0.48 & 0.48 & 0.52 & 0.49 & 0.55 & 0.81 & 0.37 \\ \hline 
$\sigma_v$ [km~s$^{-1}$] & 2.2 & 1.9 & 1.7 & 1.4 & 1.3 & 1.5 & 2.3 & 1.4 \\ \hline 
$S_\psi$ [deg] & 19.95 & 24.44 & 10.73 & 14.9 & 11.8 & 28.74 & 35.62 & 42.91 \\ \hline 
$T$ [K] & 106 & 143 & 130 & 194 & 220 & 238 & 290 & 502 \\ 
 & \textbf{503} & \textbf{517} & \textbf{843} & \textbf{1581} & \textbf{1193} & \textbf{1457} & \textbf{2322} & \textbf{2044} \\ \hline 
$x_{\rm ion}$ [$10^{-4}$] & 4.2 & 6.3 & 5.6 & 8.9 & 10.5 & 11.5 & 14.2 & 24.3 \\ 
 & \textbf{23.2} & \textbf{24.6} & \textbf{35.3} & \textbf{59.2} & \textbf{48.4} & \textbf{56.9} & \textbf{82.8} & \textbf{74.8} \\ \hline 
$A_{\rm ion}$ & 5.6 & 4.0 & 4.8 & 3.8 & 2.9 & 2.7 & 2.4 & 1.7 \\ 
 & \textbf{1.9} & \textbf{1.8} & \textbf{1.7} & \textbf{1.5} & \textbf{1.4} & \textbf{1.3} & \textbf{1.2} & \textbf{1.2} \\ \hline 
$B_{\rm sky}^{\rm DCF}$($S_v$) ~[$\mu$G] & 3.9 $\pm$ 1.5 & 1.7 $\pm$ 0.9 & 3.6 $\pm$ 2.3 & 3.2 $\pm$ 1.8 & 3.9 $\pm$ 2.8 & 1.2 $\pm$ 0.7 & 1.1 $\pm$ 0.4 & 0.4 $\pm$ 0.2 \\ 
 & \textbf{1.7} $\pm$ \textbf{0.7} & \textbf{0.9} $\pm$ \textbf{0.5} & \textbf{1.5} $\pm$ \textbf{1.0} & \textbf{1.4} $\pm$ \textbf{0.8} & \textbf{2.0} $\pm$ \textbf{1.4} & \textbf{0.6} $\pm$ \textbf{0.4} & \textbf{0.5} $\pm$ \textbf{0.2} & \textbf{0.2} $\pm$ \textbf{0.1} \\ \hline 
$B_{\rm sky}^{\rm DCF}$($\sigma_v$) ~[$\mu$G]  & 10.1 $\pm$ 4.2 & 6.8 $\pm$ 3.7 & 12.7 $\pm$ 6.1 & 9.2 $\pm$ 6.5 & 10.9 $\pm$ 8.4 &
3.2 $\pm$ 1.6 & 3.3 $\pm$ 1.2 & 1.2 $\pm$ 0.4 \\
 & \textbf{4.3} $\pm$ \textbf{1.8} & \textbf{3.5} $\pm$ \textbf{1.9} & \textbf{5.4} $\pm$ \textbf{2.6} & \textbf{3.9} $\pm$ \textbf{2.8} & \textbf{5.6} $\pm$ \textbf{4.3} & \textbf{1.6} $\pm$ \textbf{0.8} & \textbf{1.6} $\pm$ \textbf{0.6} & \textbf{0.8} $\pm$ \textbf{0.2} \\ \hline 
$B_{\rm sky}^{\rm ST}$($S_v$) ~[$\mu$G]   & 3.2 $\pm$ 1.2 & 1.4 $\pm$ 0.7 & 2.1 $\pm$ 1.4 & 1.9 $\pm$ 0.7 & 1.9 $\pm$ 0.9 & 1.1 $\pm$ 0.5 & 1.2 $\pm$ 0.4 & 0.4 $\pm$ 0.2 \\ 
 & \textbf{1.3} $\pm$ \textbf{0.5} & \textbf{0.7} $\pm$ \textbf{0.4} & \textbf{0.9} $\pm$ \textbf{0.6} & \textbf{0.8} $\pm$ \textbf{0.3} & \textbf{0.9} $\pm$ \textbf{0.5} & \textbf{0.5} $\pm$ \textbf{0.2} & \textbf{0.6} $\pm$ \textbf{0.2} & \textbf{0.3} $\pm$ \textbf{0.1} \\ \hline 
$B_{\rm sky}^{\rm ST}$($\sigma_v$) ~[$\mu$G] & 8.6 $\pm$ 2.8 & 5.8 $\pm$ 2.7 & 7.6 $\pm$ 3.3 & 5.4 $\pm$ 2.6 & 5.1 $\pm$ 2.4 & 3.0 $\pm$ 0.9 & 3.5 $\pm$ 0.9 & 1.5 $\pm$ 0.3 \\ 
 & \textbf{3.6} $\pm$ \textbf{1.2} & \textbf{3.0} $\pm$ \textbf{1.4} & \textbf{3.2} $\pm$ \textbf{1.4} & \textbf{2.3} $\pm$ \textbf{1.1} & \textbf{2.6} $\pm$ \textbf{1.2} & \textbf{1.5} $\pm$ \textbf{0.5} & \textbf{1.7} $\pm$ \textbf{0.4} & \textbf{0.9} $\pm$ \textbf{0.2} \\ \hline 
$\Gamma_{\rm in,S16}$~[$ 10^{-9}$~s$^{-1}$] & 6.4 & 6.1 & 6.3 & 6.1 & 5.9 & 5.9 & 5.9 & 6.0 \\ 
 & \textbf{6.0} & \textbf{6.0} & \textbf{6.0} & \textbf{6.0} & \textbf{6.1} & \textbf{6.1} & \textbf{6.1} & \textbf{6.2} \\ \hline 
$\kappa_\parallel$~[$ 10^{28}$~cm$^2$~s$^{-1}$] & 6.9 $\pm$ 1.2 & 5.3 $\pm$ 0.6 & 6.2 $\pm$ 1.5 & 5.2 $\pm$ 1.3 & 4.4 $\pm$ 0.4 & 4.2 $\pm$ 0.3 & 4.0 $\pm$ 0.3 & 3.5 $\pm$ 0.07 \\ 
 & \textbf{3.7} $\pm$ \textbf{0.2} & \textbf{3.6} $\pm$ \textbf{0.1} & \textbf{3.6} $\pm$ \textbf{0.2} & \textbf{3.4} $\pm$ \textbf{0.3} & \textbf{3.4} $\pm$ \textbf{0.08} & \textbf{3.4} $\pm$ \textbf{0.04} & \textbf{3.4} $\pm$ \textbf{0.1} & \textbf{3.4} $\pm$ \textbf{0.02} \\ \hline 
$M_A^{DCF}$ & 0.99 $\pm$ 0.31 & 1.21 $\pm$ 0.54 & 0.53 $\pm$ 0.22 & 0.74 $\pm$ 0.45 & 0.58 $\pm$ 0.46 & 1.42 $\pm$ 0.56 & 1.76 $\pm$ 0.53 & 2.12 $\pm$ 0.47 \\ \hline 
$M_A^{ST}$ & 1.17 $\pm$ 0.17 & 1.28 $\pm$ 0.27 & 0.85 $\pm$ 0.17 & 0.97 $\pm$ 0.31 & 0.85 $\pm$ 0.32 & 1.39 $\pm$ 0.28 & 1.56 $\pm$ 0.25 & 1.72 $\pm$ 0.2 \\ \hline 
$M_S$ ($S_v$) & 1.35 $\pm$ 0.49 & 0.72 $\pm$ 0.37 & 0.72 $\pm$ 0.63 & 0.68 $\pm$ 0.22 & 0.6 $\pm$ 0.18 & 0.62 $\pm$ 0.26 & 0.84 $\pm$ 0.36 & 0.29 $\pm$ 0.11 \\ 
 & \textbf{0.67} $\pm$ \textbf{0.28} & \textbf{0.39} $\pm$ \textbf{0.21} & \textbf{0.31} $\pm$ \textbf{0.24} & \textbf{0.27} $\pm$ \textbf{0.1} & \textbf{0.27} $\pm$ \textbf{0.09} & \textbf{0.25} $\pm$ \textbf{0.1} & \textbf{0.3} $\pm$ \textbf{0.14} & \textbf{0.15} $\pm$ \textbf{0.05} \\ \hline 
$M_S$ ($\sigma_v$) & 3.83 $\pm$ 1.01 & 2.93 $\pm$ 1.3 & 2.57 $\pm$ 1.49 & 1.96 $\pm$ 0.48 & 1.63 $\pm$ 0.3 & 1.77 $\pm$ 0.33 & 2.39 $\pm$ 0.58 & 1.09 $\pm$ 0.22 \\ 
 & \textbf{1.92} $\pm$ \textbf{0.63} & \textbf{1.6} $\pm$ \textbf{0.77} & \textbf{1.14} $\pm$ \textbf{0.58} & \textbf{0.82} $\pm$ \textbf{0.34} & \textbf{0.74} $\pm$ \textbf{0.18} & \textbf{0.74} $\pm$ \textbf{0.17} & \textbf{0.86} $\pm$ \textbf{0.22} & \textbf{0.54} $\pm$ \textbf{0.11} \\ \hline 
\end{tabular}
}
\renewcommand{\arraystretch}{1}
\label{tab:RetISMprop}
\end{table*}
% ********************************************************
\begin{table*}[!ht]
\caption{ISM properties along the Eridu filament, for the fiber (regular font) or shell (bold font) geometry. The values and errors respectively give the mean value in each ${\sim}5$~pc-wide square and the rms dispersion at 0.5--0.7~pc scale within each square. The Galactic coordinates refer to the square centre.}
\centering
\scalebox{0.85}{
\renewcommand{\arraystretch}{2}
\begin{tabular}{cccccccc}
\hline
long. [deg] & 194.05 & 196.73 & 199.49 & 203.1 & 205.33 & 206.79 & 224.21 \\ 
lat. [deg] & -56.77 & -57.76 & -59.73 & -61.29 & -62.52 & -64.22 & -62.46 \\ \hline 
$N_{\rm HI}$ [$ 10^{20}$~cm$^{-2}$] & 2.0 & 1.8 & 1.7 & 2.6 & 1.9 & 2.0 & 0.9 \\ \hline 
LOS depth [pc]  & 8.1 $\pm$ 0.8 & 8.4 $\pm$ 0.8 & 11.4 $\pm$ 1.1 & 8.8 $\pm$ 0.9 & 16.1 $\pm$ 1.6 & 9.2 $\pm$ 0.9 & 6.7 $\pm$ 0.7 \\ 
 & \textbf{30.0} $\pm$ \textbf{3.0} & \textbf{30.0} $\pm$ \textbf{3.0} & \textbf{30.0} $\pm$ \textbf{3.0} & \textbf{30.0} $\pm$ \textbf{3.0} & \textbf{30.0} $\pm$ \textbf{3.0} & \textbf{30.0} $\pm$ \textbf{3.0} & \textbf{30.0} $\pm$ \textbf{3.0} \\ \hline 
$n_{\rm HI}$ [cm$^{-3}$] & 9 $\pm$ 4 & 7 $\pm$ 2 & 5 $\pm$ 2 & 10 $\pm$ 1.7 & 5 $\pm$ 1 & 7 $\pm$ 1 & 5 $\pm$ 1 \\ 
& \textbf{2} $\pm$ \textbf{0.8} & \textbf{2} $\pm$ \textbf{0.5} & \textbf{2} $\pm$ \textbf{0.7} & \textbf{3} $\pm$ \textbf{0.5} & \textbf{2} $\pm$ \textbf{0.6} & \textbf{2} $\pm$ \textbf{0.2} & \textbf{1} $\pm$ \textbf{0.2} \\ \hline 

$S_v$ [km~s$^{-1}$] & 1.11 & 1.49 & 1.27 & 1.0 & 1.01 & 1.7 & 0.47 \\ \hline 
$\sigma_v$ [km~s$^{-1}$] & 3.2 & 3.3 & 3.7 & 3.0 & 2.8 & 3.3 & 3.4 \\ \hline 
$S_\psi$ [deg] & 11.62 & 14.32 & 18.15 & 9.41 & 15.05 & 19.01 &  \\ \hline 
$T$ [K]  & 488 & 440 & 789 & 295 & 1102 & 408 & 845 \\ 
 & \textbf{2847} & \textbf{2867} & \textbf{3202} & \textbf{1592} & \textbf{2673} & \textbf{2317} & \textbf{5868} \\ \hline 
$x_{\rm ion}$ [$10^{-4}$] & 22.2 & 21.4 & 34.6 & 14.6 & 45.3 & 20.1 & 36.9 \\ 
 & \textbf{101.2} & \textbf{98.2} & \textbf{108.9} & \textbf{61.4} & \textbf{94.4} & \textbf{82.7} & \textbf{200.4} \\ \hline 
$A_{\rm ion}$ & 2.0 & 1.9 & 1.6 & 2.3 & 1.5 & 1.9 & 1.5 \\ 
 & \textbf{1.2} & \textbf{1.2} & \textbf{1.2} & \textbf{1.3} & \textbf{1.2} & \textbf{1.2} & \textbf{1.1} \\ \hline 
$B_{\rm sky}^{\rm DCF}$($S_v$) ~[$\mu$G] & 4.6 $\pm$ 2.8 & 4.9 $\pm$ 2.3 & 4.6 $\pm$ 5.4 & 6.4 $\pm$ 4.0 & 2.2 $\pm$ 0.8 & 4.1 $\pm$ 1.6 &  \\ 
 & \textbf{2.4} $\pm$ \textbf{1.5} & \textbf{2.6} $\pm$ \textbf{1.2} & \textbf{2.9} $\pm$ \textbf{3.3} & \textbf{3.4} $\pm$ \textbf{2.2} & \textbf{1.6} $\pm$ \textbf{0.6} & \textbf{2.3} $\pm$ \textbf{0.9} & \\ \hline 
$B_{\rm sky}^{\rm DCF}$($\sigma_v$) ~[$\mu$G] & 10.7 $\pm$ 4.6 & 9.0 $\pm$ 4.6 & 9.1 $\pm$ 5.5 & 15.4 $\pm$ 8.6 & 5.9 $\pm$ 2.2 & 7.1 $\pm$ 3.3 &  \\ 
 & \textbf{5.3} $\pm$ \textbf{2.3} & \textbf{4.8} $\pm$ \textbf{2.5} & \textbf{5.6} $\pm$ \textbf{3.4} & \textbf{8.4} $\pm$ \textbf{4.7} & \textbf{3.9} $\pm$ \textbf{1.5} & \textbf{4.0} $\pm$ \textbf{1.8} &  \\ \hline 
$B_{\rm sky}^{\rm ST}$($S_v$) ~[$\mu$G] & 2.8 $\pm$ 1.5 & 3.2 $\pm$ 1.1 & 2.5 $\pm$ 2.0 & 3.1 $\pm$ 1.2 & 1.5 $\pm$ 0.4 & 3.1 $\pm$ 0.7 &  \\ 
 & \textbf{1.4} $\pm$ \textbf{0.8} & \textbf{1.7} $\pm$ \textbf{0.6} & \textbf{1.6} $\pm$ \textbf{1.2} & \textbf{1.7} $\pm$ \textbf{0.6} & \textbf{1.1} $\pm$ \textbf{0.3} & \textbf{1.7} $\pm$ \textbf{0.4} & \\ \hline 
$B_{\rm sky}^{\rm ST}$($\sigma_v$) ~[$\mu$G] & 7.3 $\pm$ 2.6 & 6.3 $\pm$ 2.4 & 6.0 $\pm$ 2.2 & 8.5 $\pm$ 2.5 & 4.4 $\pm$ 1.1 & 5.7 $\pm$ 1.5 & \\ 
 & \textbf{3.6} $\pm$ \textbf{1.3} & \textbf{3.4} $\pm$ \textbf{1.3} & \textbf{3.7} $\pm$ \textbf{1.3} & \textbf{4.6} $\pm$ \textbf{1.4} & \textbf{2.9} $\pm$ \textbf{0.8} & \textbf{3.2} $\pm$ \textbf{0.8} & \\ \hline 
$\Gamma_{ \rm in,S16}$ ~[$ 10^{-9}$~s$^{-1}$] & 6.0 & 5.9 & 6.0 & 5.9 & 6.1 & 5.9 & 6.1 \\ 
 & \textbf{5.9} & \textbf{6.0} & \textbf{5.9} & \textbf{6.2} & \textbf{6.0} & \textbf{6.2} & \textbf{4.9} \\ \hline 
$\kappa_\parallel$ ~[$ 10^{28}$~cm$^2$~s$^{-1}$] & 3.7 $\pm$ 0.3 & 3.6 $\pm$ 0.3 & 3.5 $\pm$ 0.1 & 3.9 $\pm$ 0.2 & 3.4 $\pm$ 0.08 & 3.6 $\pm$ 0.07 & 3.4 $\pm$ 0.05 \\ 
 & \textbf{3.3} $\pm$ \textbf{0.4} & \textbf{3.3} $\pm$ \textbf{0.3} & \textbf{3.3} $\pm$ \textbf{0.2} & \textbf{3.4} $\pm$ \textbf{0.03} & \textbf{3.3} $\pm$ \textbf{0.3} & \textbf{3.4} $\pm$ \textbf{0.06} & \textbf{2.7} $\pm$ \textbf{0.4} \\ \hline 
$M_A^{DCF}$ & 0.57 $\pm$ 0.19 & 0.71 $\pm$ 0.38 & 0.9 $\pm$ 0.7 & 0.46 $\pm$ 0.2 & 0.74 $\pm$ 0.25 & 0.94 $\pm$ 0.37 & \\ \hline 
$M_A^{ST}$ & 0.89 $\pm$ 0.14 & 0.97 $\pm$ 0.25 & 1.05 $\pm$ 0.4 & 0.79 $\pm$ 0.18 & 1.01 $\pm$ 0.17 & 1.13 $\pm$ 0.22 & \\ \hline 
$M_S$ ($S_v$) & 1.02 $\pm$ 0.53 & 1.3 $\pm$ 0.51 & 0.88 $\pm$ 0.5 & 1.03 $\pm$ 0.27 & 0.57 $\pm$ 0.14 & 1.48 $\pm$ 0.33 & 0.85 $\pm$ 0.16 \\ 
 & \textbf{0.41} $\pm$ \textbf{0.23} & \textbf{0.52} $\pm$ \textbf{0.23} & \textbf{0.44} $\pm$ \textbf{0.26} & \textbf{0.45} $\pm$ \textbf{0.12} & \textbf{0.36} $\pm$ \textbf{0.08} & \textbf{0.62} $\pm$ \textbf{0.14} & \textbf{0.31} $\pm$ \textbf{0.03} \\ \hline 
$M_S$ ($\sigma_v$) & 2.93 $\pm$ 1.02 & 2.85 $\pm$ 0.97 & 2.47 $\pm$ 0.69 & 3.05 $\pm$ 0.54 & 1.64 $\pm$ 0.42 & 2.89 $\pm$ 0.6 & 1.7 $\pm$ 0.31 \\ 
 & \textbf{1.18} $\pm$ \textbf{0.42} & \textbf{1.14} $\pm$ \textbf{0.42} & \textbf{1.22} $\pm$ \textbf{0.37} & \textbf{1.34} $\pm$  \textbf{0.27} & \textbf{1.02} $\pm$ \textbf{0.25} & \textbf{1.21} $\pm$ \textbf{0.25} & \textbf{0.61} $\pm$ \textbf{0.06} \\ 
 \hline 
\end{tabular}}
\renewcommand{\arraystretch}{1}
\label{tab:EriISMprop}
\end{table*}

\end{document}